\documentclass{aa}  

\usepackage{amsmath}
\usepackage{amsfonts}
\usepackage{amssymb}
\usepackage{graphicx}
\usepackage{bm}
\usepackage{gensymb}
\usepackage[utf8]{inputenc}
\usepackage{subcaption}
\usepackage{fixltx2e}
\usepackage{wrapfig}
\usepackage{epstopdf}
\usepackage{dirtytalk}
\usepackage{comment}
\AppendGraphicsExtensions{.tiff}
\usepackage{txfonts}
\begin{document}

   \title{Effect of the solar wind density on the evolution of normal and inverse coronal mass ejections}


   \author{S. Hosteaux,
          E. Chané
                  ,
                  S. Poedts
          }

   \institute{Centre for mathematical Plasma-Astrophysics (CmPA), KU Leuven,
              Celestijnenlaan 200B, 
                           B-3001 Leuven
             }


 
  \abstract
   {The evolution of magnetised coronal mass
ejections (CMEs) and their interaction with the background solar wind leading to deflection, deformation, and erosion is still largely unclear as there is very little observational data available. Even so, this evolution is very important for the geo-effectiveness of CMEs.}
   {We investigate the evolution of both normal and inverse CMEs ejected at different initial velocities, and observe the effect of the background wind density and their magnetic polarity on their evolution up to 1~AU.}
   {We performed 2.5D (axisymmetric) simulations by solving the magnetohydrodynamic (MHD) equations on a radially stretched grid, employing a block-based adaptive mesh refinement (AMR) scheme based on a density threshold to achieve high resolution following the evolution of the magnetic clouds and the leading bow shocks. All the simulations discussed in the present paper were performed using the same initial grid and numerical methods.}
   {The polarity of the internal magnetic field of the CME has a substantial effect on its propagation velocity and on its deformation and erosion during its evolution towards Earth. We quantified the effects of the polarity of the internal magnetic field of the CMEs and of the density of the background solar wind on the arrival times of the shock front and the magnetic cloud. We determined the positions and propagation velocities of the magnetic clouds and thus also the stand-off distance of the leading shock fronts (i.e.\ the thickness of the magnetic sheath region) and the deformation and erosion of the magnetic clouds during their evolution from the Sun to the Earth. Inverse CMEs were found to be faster than normal CMEs ejected in the same initial conditions, but with smaller stand-off distances. They also have a higher magnetic cloud length, opening angle, and mass. Synthetic satellite time series showed that the shock magnitude is not affected by the polarity of the CME. However, the density peak of the magnetic cloud is dependent on the polarity and, in case of inverse CMEs, also on the background wind density. The magnitude of the z-component of the magnetic field was not influenced by either the polarity or the wind density.}
   {}


   \titlerunning{}
   \authorrunning{S. Hosteaux et al.}
   \maketitle
   
%

\section{Introduction}

Coronal mass ejections (CMEs) are violent, large-scale eruptions of plasma and magnetic field originating from the solar corona. Together with coronal interaction regions, CMEs are the main causes of geomagnetic storms. Whether a geomagnetic storm occurs or not depends, among other parameters, on the sign and magnitude of the $z$-component of the internal magnetic field of the CME (perpendicular to the equatorial plane) \citep{Tsurutani1986}. If the polarity of the CME magnetic field is opposite to that of the magnetic field the magnetosphere of the Earth, magnetic reconnection occurs at the magnetopause, allowing  more charged particles to enter the magnetosphere. The most important properties of CMEs that determine the intensity of the geomagnetic storm they cause are their velocity, their magnetic field (strength and orientation), and to a lesser extent CME mass. The velocity of the CMEs at 1~AU can range from relatively slow (about $250\;$km/s) to extreme speeds of about $2200\;$km/s \citep{Russell2013,Liu2014}, while the mass of CMEs ranges from $10^{11}$~kg to $10^{13}$~kg \citep{Hudson1996,Jackson1985}. CMEs that contain magnetic clouds have a magnetic field strength typically between 15 and 30~nT at 1~AU \citep{Lepping1990}.

The structure of most fast interplanetary coronal mass ejections (ICMEs) can generally be divided into three main components: a leading shock front, followed by a sheath region, followed in turn by a magnetic cloud \citep{Illing1985}. If the speed of the CME exceeds the local speed of the fast magnetosonic wave in the solar wind frame, a shock will develop in front of the ejected material, forming a discontinuity in plasma properties such as density, pressure, and velocity. Strong interplanetary shock waves that impact  the Earth's magnetosphere are often related to intense  (Dst < -100~nT) or moderate (50~nT $\leq$ Dst < -100~nT) geomagnetic disturbances \citep{Echer2004, Huttunen2004, Lugaz2016}. 

Furthermore, the propagating CME shocks are able to drive solar energetic particles (SEP) as the particles are accelerated diffusively at the shock front, causing gradual SEP events \citep{Li2003, Reames2013}. The turbulent region immediately behind the shock is called the CME sheath. Observations show that this region has larger magnetic field variations, higher temperatures, and higher densities compared to the following magnetic cloud. The thickness of the sheath region is also called the stand-off distance of the ICME shock front. The compression of the plasma inside the sheath region makes it a very suitable location for magnetic reconnection \citep{Kilpua2017}. 

The sheath region of the ICME is followed by a magnetic cloud, which is a large closed field structure of increased magnetic field strength and below average temperature, and whose magnetic field
direction rotates smoothly. Not all observed ICMEs display this three-part structure, however, as not all ICMEs develop a shock wave in front of them and approximately one-third \citep{Gosling1990} to one-half \citep{Cane1997} of all observed ICMEs show signatures of a magnetic cloud, though this might be due to the fact that most observations are made by one spacecraft on a single track through the ICME, and thus might monitor only part of the ICME.\par

Though recent progress has increased our understanding of the initial onset phase of a CME eruption, there is no consensus on what mechanism(s) initiate(s) a CME. Several self-consistent CME onset models have been and are still being extensively researched, most based-on the fact that the energy input of the eruption mainly comes from the coronal magnetic field \citep{Mikic1994,vanderHolst2005, Hosteaux2018}. Another set of MHD models focuses more on the propagation of ICMEs and their interaction with the background solar wind such as work by \cite{Chane2005,Chane2006,Chane2008,Jacobs2005} and \cite{vanderHolst2005}, which used a relatively simple magnetised high density--pressure blob model. Understanding and predicting how the structure and properties of a CME and its preceding shock evolves from its ejection to 1~AU is one of the main goals of space weather research, for which global MHD modelling has proven to be a useful tool \citep{Shiota2016, Lionello2013, Lugaz2011, Pomoell2018, Zhou2017, Temmer2011, Mao2017}. For example, \cite{Zhou2017} used a 3D MHD CME model to analyse the propagation characteristics of CMEs launched at different latitudes. These authors found that the arrival time of the shock is dependent on whether or not the CME is launched at the same side of the heliospheric current sheet (HCS) as the Earth, and that ICMEs deflect towards the HCS. Also, \cite{Temmer2011} used the 3D MHD heliospheric wind and non-magnetised CME evolution model ENLIL \citep{Odstrcil2004} to study the influence of the solar wind on the propagation of ICMEs. Axisymmetric (2.5D) simulations using a model that artificially imposes a flux rope on a background wind were performed by \cite{Savani2012} to investigate the heliocentric distance dependence of the stand-off  distance for CMEs with different initial conditions. \cite{Torok2018} were able to reproduce an MHD simulation of the extreme Bastille Day event using a realistic initiation mechanism. This was the first time that MHD modelling managed to reproduce impulsive eruptions propagating to 1AU starting from stable magnetic configurations. A study by \cite{Jin2017} performed data-driven MHD simulations using the Gibson-Low flux rope model in the interest of developing MHD models as CME forecasting tools. \par

To better understand how an ICME evolves as it propagates, it is important to distinguish between the effects of the CME initialisation parameters (e.g.\ initial speed, magnetic polarity) and the effect of the background wind.
In the present study, we are interested in both aspects and their effect on the evolution of an ICME during its propagation to 1~AU. Here we focus on the influence of the surrounding background wind speed and density, the initial CME velocity and the magnetic field polarity of the CME. For this reason, and to have full control over all aspects of the initial CME that is leaving the solar corona (size, total mass, speed, magnetic field strength, and polarity, etc.), we use an  initial CME set-up similar to that used by \cite{Chane2005}. Our simple density-driven model superimposes a magnetised high density plasma blob on a background solar wind, neglecting the initiation phase of the CME but providing total control of the initial state of the ejection. This magnetised density-driven model is used to perform high resolution simulations of ICMEs propagating through different background solar winds up to 1~AU. We performed seventeen 2.5D MHD simulations to study the dependence of the ICME evolution on the density of the background wind, the initial CME speed and the initial polarity of the internal magnetic field of the CME. Below, we  discuss the background solar wind model used in these simulations, and in the next section we discuss the simple (but magnetised) CME model used and the important initial parameters. We examine the kinematics of both shock and magnetic cloud in section \ref{subsec:Kinematics}. The deformation of the magnetic cloud is discussed in section \ref{subsec:deformation}. Finally, synthetic satellite data of the ICME parameters at L1 is investigated and discussed in section \ref{subsec:sat}.

\section{Background solar wind model}

All solar wind and CME simulations discussed in the present paper were performed by using MPI-AMRVAC \citep{Porth2014, Xia2018} to numerically solve the MHD equations on a spherical logarithmically stretched grid. We performed our simulations in 2.5D, meaning that the MHD equations are solved on a 2D mesh, so the plasma quantities do not have any $\phi$ dependence, but all three vector components (radial distance $r$, polar angle $\theta$, and azimuthal angle $\phi$ in spherical geometry) of the velocity and magnetic fields are included in the calculations.  \cite{Jacobs2007} proved that fully 3D simulations of CMEs are very well approximated by carefully set-up 2.5D simulations (if  the 2.5D CMEs have the same momentum density as the 3D CMEs), while being two orders of magnitude more CPU efficient. The numerical domain comprises $[1,216]\;$R$_{\odot}$ in the radial ($r$-) direction and $[0,\pi]\;$radians in the latitudinal ($\theta$-) direction. The base grid consists of $300 \times 220$ cells (logarithmically stretched) and a Lohner block-based adaptive mesh refinement scheme is used, refining or coarsening the blocks that form the grid once a chosen density gradient threshold is reached. The blocks are refined following the ejection up to a maximum of four levels, meaning three levels on top of the base grid. All simulations were performed with the same base grid and the same numerical methods, so we can attribute the differences between the different simulations to the different initialisation parameters.\par

We consider solar minimum conditions. In addition to the gravitational force, an extra empirical heating and cooling source term, as described in \cite{Manchester2004}, is implemented to achieve a realistic bi-modal fast and slow solar wind:

\begin{equation}
Q=\rho q_0 (T_0 - T)\exp\left[-\frac{(r-1)^2}{\sigma^2_0}\right] \hspace{0.1cm},
\end{equation}with $q_0$ the volumetric heating amplitude, $T_0$ the target temperature, and $\sigma_0$ the heating scale height. Both the target temperature and the heating scale height are latitude-dependent. In the equatorial region, $T_0 = 1.5\times 10^6\;$K and $\sigma_0 = 4.5\;$R$_{\odot}$ up to a certain critical angle. Polewards from the equator $T_0$ becomes $2.63\times 10^6\;$K and $\sigma_0$ increases as $4.5 \left[2-\sin^2(\theta)/\sin^2(\theta_0)\right]\;$R$_{\odot}$. This critical angle is dependent on the radial distance: for $r\leq7R_{\odot}$, $\sin^2 \theta=\sin^2(17.5\degree)+\cos^2(17.5\degree)(r/R_\odot-1)/8$, while for $r>7R_{\odot}$ this becomes $\sin^2 \theta=\sin^2(61.5\degree)+\cos^2(61.5\degree)(r/R_\odot-7)/40$. Furthermore, a magnetic dipole field with a strength of $2.2\;$G at the poles is imposed as a boundary condition, as explained in the next paragraph. \par

\begin{figure}
   \centering
   \includegraphics[width=\hsize]{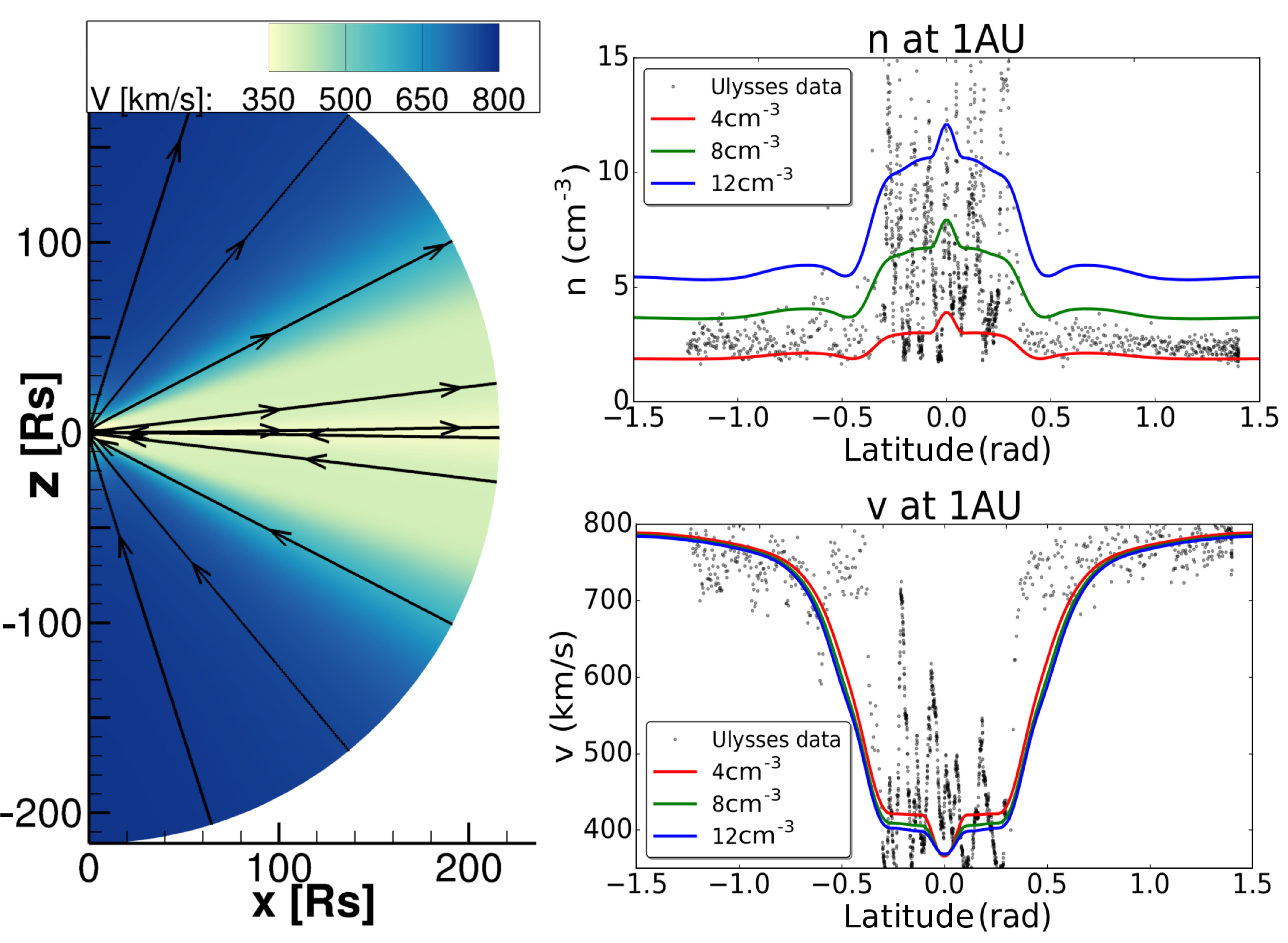}
      \caption{Left column: Plasma speed and magnetic field lines for a medium density wind. Right column: Density and velocity profiles at 1~AU in the $\theta$-direction for three winds with different densities. They are superposed on Ulysses measurements of the plasma density that have been normalised to 1~AU. The axes denote the distance from the centre of the Sun in Cartesian coordinates in solar radii.
      }
         \label{fig:wind}
   \end{figure}

At the inner boundary, the number density and temperature are fixed to $10^8\;$cm\textsuperscript{-3} and $1.5\times10^6\;$K, respectively. In addition, $r^2B_r$ is also fixed to enforce the dipole field and $B_\theta$ decreases as a dipole. A dead-zone of $\pm 45\degree$ around the equator is imposed, where at the boundary $v_r=0$, $v_\theta=0$, and $B_\phi$ is extrapolated. Outside the dead-zone $r^2 \rho v_r$ and $r^3(\rho v_r v_\phi-B_r B_\phi)$ are extrapolated and $v_\theta$ is chosen so that $\vec{B}$ and $\vec{v}$ are parallel. At the outer boundary $r^2 \rho$, $r^2 \rho v_r$, $\rho v_\theta$, $r v_\phi$, $r^2 B_r$, $B_\theta$, $r B_\phi$, and $T$ are all extrapolated. The left part of figure~\ref{fig:wind} shows the solar wind speed for the whole domain together with some selected magnetic field lines. We let each background wind simulation run for 1000 hours to ensure a fully relaxed solar wind was reached. This relaxed state acts as the initial state ($t=0\;$h) upon which the CMEs were imposed. \par

In order to study the effect of the background solar wind density on the propagation of the CMEs, three different background winds were used in our simulations. The upper panel of the right column of figure~\ref{fig:wind} shows the density profiles in the polar direction at 1~AU of the three winds together with observational data from Ulysses from November 1994 to January 1995 (close to solar minimum). The Ulysses data shows a large variety in solar wind density, validating our choice of background winds. It can be seen in the figure that three different solar winds were used, namely a wind with a density of 4 cm\textsuperscript{-3} on the equatorial plane at 1~AU, another wind with a density of 8 cm\textsuperscript{-3}, and finally one with a density of 12 cm\textsuperscript{-3}, which are referred to as the low, medium, and high density solar wind, respectively. The different background winds were obtained by changing the density of the inner boundary. The bottom panel of figure \ref{fig:wind} shows the velocity profile at 1~AU. The velocity difference between the different simulated background winds is negligible and they match  the Ulysses observations of the solar wind fairly well. 

\section{Blob CME model}

   \begin{figure}
   \centering
   \includegraphics[width=\hsize]{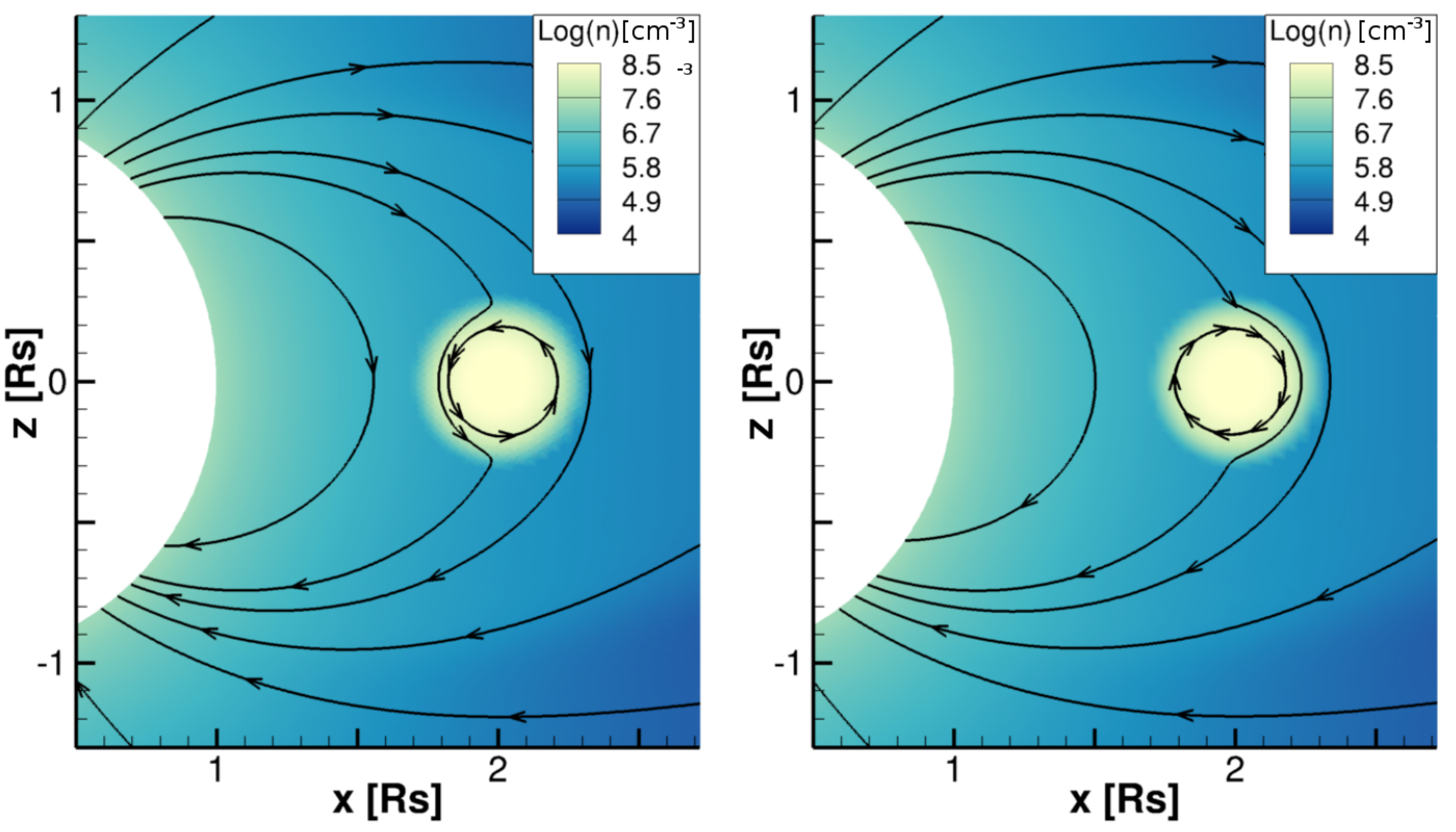}
      \caption{Selected magnetic field lines and density contours at $t=0$ for a normal CME (left) and an inverse CME (right).
              }
         \label{fig:initial_blob}
   \end{figure}

Since the focus of the present study is the evolution of the CMEs and not   their initiation, a very simple model was used that omits the onset of the CME. A magnetised sphere of high density and high pressure is superimposed on a relaxed solar wind and is given an initial radial velocity. The density-velocity-pressure profile of the blob is given by \cite{Chane2005} and \cite{Jacobs2005}: 

\begin{equation}
w=\frac{w_{cme}}{2}\left(1-\cos\,\pi\frac{d_{cme}-d}{d_{cme}}\right) \hspace{0.1cm}.
\end{equation}Here $w$ represents either density, pressure, or radial velocity; $w_{cme}$ is the maximum value of these parameters inside the sphere; $d_{cme}$ corresponds to the radius of the blob; and $d$ denotes the distance to the centre of the blob. This profile ensures a smooth transition between the perturbed region of the CME and the ambient solar wind. Using the same method described in \cite{Jacobs2007}, guaranteeing a 2.5D evolution very similar to a full 3D evolution, the density was chosen so that the mass of the torus is equivalent to that of a 3D sphere with a total mass of $1.267 \times 10^{16}\;$g. This yields a total torus mass of $2.74 \times 10^{17}\;$g and a torus density 25 times higher than that of the solar surface. Because the initial CME temperature is uniform, the high ratio between CME and solar wind density results in an extremely rapid expansion of the CME due to the pressure gradient between the magnetic cloud and the background wind. The centre of the CME was chosen to be at 2.0~R$_{\odot}$ on the equator with its radius being 0.29~R$_{\odot}$.

Following \cite{Chane2006}, the poloidal components ($R=r\cos{\theta}$ and $Z=r\sin{\theta}$) of the initial CME magnetic field can be written as

\begin{equation}
B_{R}=-\frac{1}{R}\frac{\partial \psi}{\partial Z},\hspace{1cm} B_{Z}=\frac{1}{R}\frac{\partial \psi}{\partial R} \hspace{0.1cm},
\end{equation}where $\psi$ denotes the magnetic flux. This flux is chosen so that it connects smoothly to the background wind:

\begin{equation}
\psi = \psi_{1} \left( d-\frac{d_{cme}}{2\pi}\sin\frac{2\pi d}{d_{cme}} \right) \hspace{0.1cm}. 
\end{equation}The sign of the constant $\psi_1$ determines the polarity of the internal magnetic field of the CME. If $\psi_1$ is negative, the initial magnetic field polarity of the CME is the same as that of the background solar coronal magnetic field resulting in what is known as an inverse CME. If it is positive, the initial magnetic field polarity of the CME is the opposite than that of the background solar wind and the corona. This results in a  normal CME. The terminology inverse and normal CMEs was introduced by \cite{Low2002}. The maximum value of the magnetic field strength in the initial blob configuration is approximately 2~mT. Figure~\ref{fig:initial_blob} shows the two different magnetic configurations that are  superposed on the background solar corona (of which a few field lines are also shown). \par

Simulations with three different background solar winds and three different initial CME velocities ($400\;$km/s, $800\;$km/s and $1200\;$km/s) were performed for both normal and inverse CMEs. Unfortunately, the high velocity inverse CME simulation superposed on a low density background wind crashed due to numerical errors.  Apparently, this  initial velocity is too high for the low density wind case, in spite of the chosen smooth profiles. We decided not to try to resolve the issue with different numerical techniques or different initialisation procedures, which would have taken considerable computational resources;  we would have had to redo  the other 17 simulations as well.

\section{Results on ICME evolution}

\begin{figure*}[!ht]
   \centering
   \includegraphics[width=0.95\hsize]{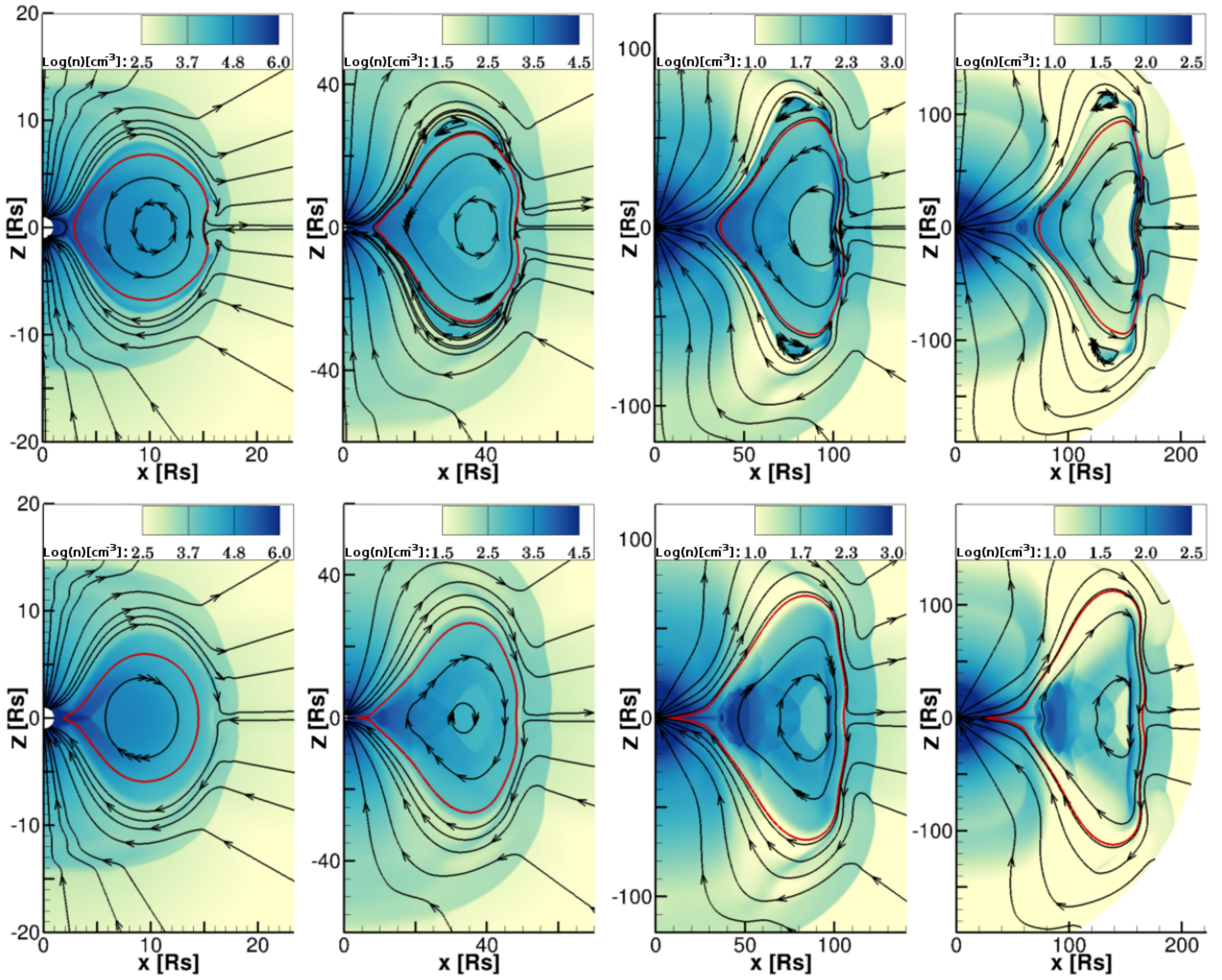}
      \caption{(from
left to right) Snapshots of the logarithmic number density (colour-coded) and selected magnetic field lines (in black) at $t=2.5\;$h, $t=10\;$h, $t=30\;$h, and $t=40\;$h after ejection  for a normal CME (upper panels) and an inverse CME (lower panels). The red line depicts the boundary of the magnetic cloud of the CME in both figures.
              }
         \label{fig:together}
   \end{figure*}

To perform an analysis of the properties of a magnetic cloud we need to know its location and size. We define the magnetic cloud of a CME in the simulations as the region or volume inside the largest closed magnetic structure in the numerical domain (i.e.\ inside the separatrix). The evolution of both a normal and inverse CME, both ejected at an initial velocity of $800\;$km/s in a medium density background wind, are illustrated by the snapshots shown in Figure~\ref{fig:together}. The closed red structure in the  panels represents the separatrix determining the magnetic cloud. The polarity of the CMEs has a substantial effect on the evolution of its shape (i.e.\ its deformation), which is due to different magnetic reconnection processes that occur during their propagation in the magnetised background wind. 
As seen in Figure~\ref{fig:initial_blob}, a normal CME has an opposite polarity with respect to the background wind at the front of the CME. This implies that a current sheet will form there, and thus magnetic reconnection will occur at the front side of the ejecta.

\begin{figure}
  \begin{center}
    \includegraphics[width=\hsize]{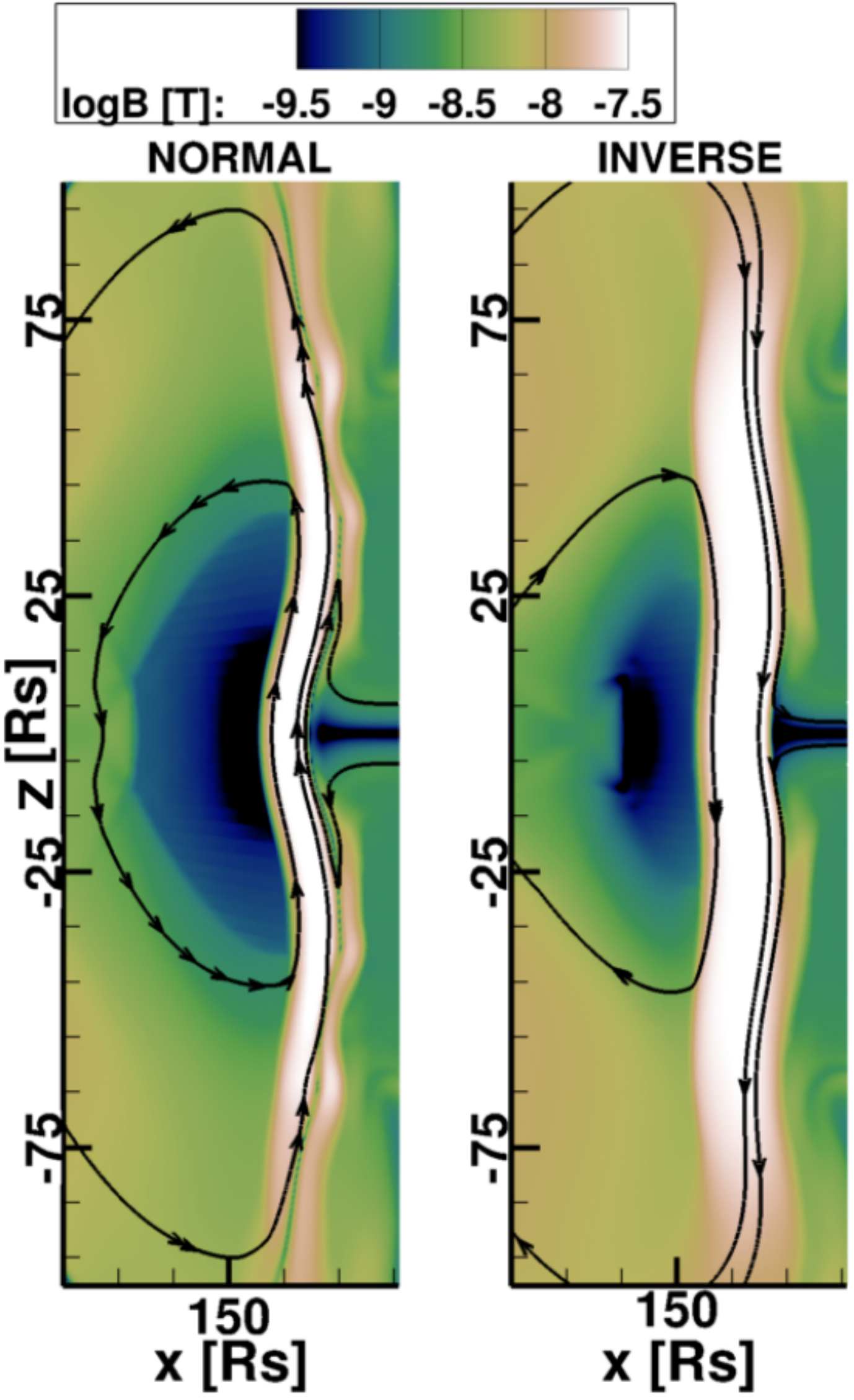}
  \end{center}
  \caption{Magnified view of the fronts of two CMEs at t=45\;h, both at an initial velocity of $800\;$km/s ejected in a medium density background wind. The CME in the left panel has a normal polarity, while the CME in the right panel has an inverse polarity.}\label{fig:front}
\end{figure}

For an inverse CME, however, the situation is reversed. The magnetic field of such a CME has the same polarity as that of the wind surrounding the CME, and hence magnetic reconnection occurs behind the CME instead of in front of it. As a result, the magnetic cloud of the normal CME appears to have a more tumultuous front than the inverse CME, with several small closed magnetic structures of increased density formed along the separatrix due to tearing instabilities occurring in the current sheet. Two larger structures are formed both on top and below the CME, with opposite polarity in comparison to the polarity of the magnetic cloud itself. Inverse CMEs, however, seem very smooth at the front of the separatrix. There are no reconnection processes occurring here, only the magnetic cloud pushing against the surrounding magnetic field. Due to reconnection processes at their back, inverse CMEs have an elongated tail end when compared to their normal counterpart, where a continuous generation of small-scale structures occurs that merge with the magnetic cloud. The difference between the front of a normal CME and the front of an inverse CME can be seen in Figure~\ref{fig:front}, where the thin and dimpled front magnetic sheet of the normal CME is in great contrast with the smooth and thick front sheet of the inverse CME.\par

\subsection{Kinematics of shocks and magnetic clouds}\label{subsec:Kinematics}

\begin{figure*}[!htb]
{\includegraphics[width=0.495\linewidth]{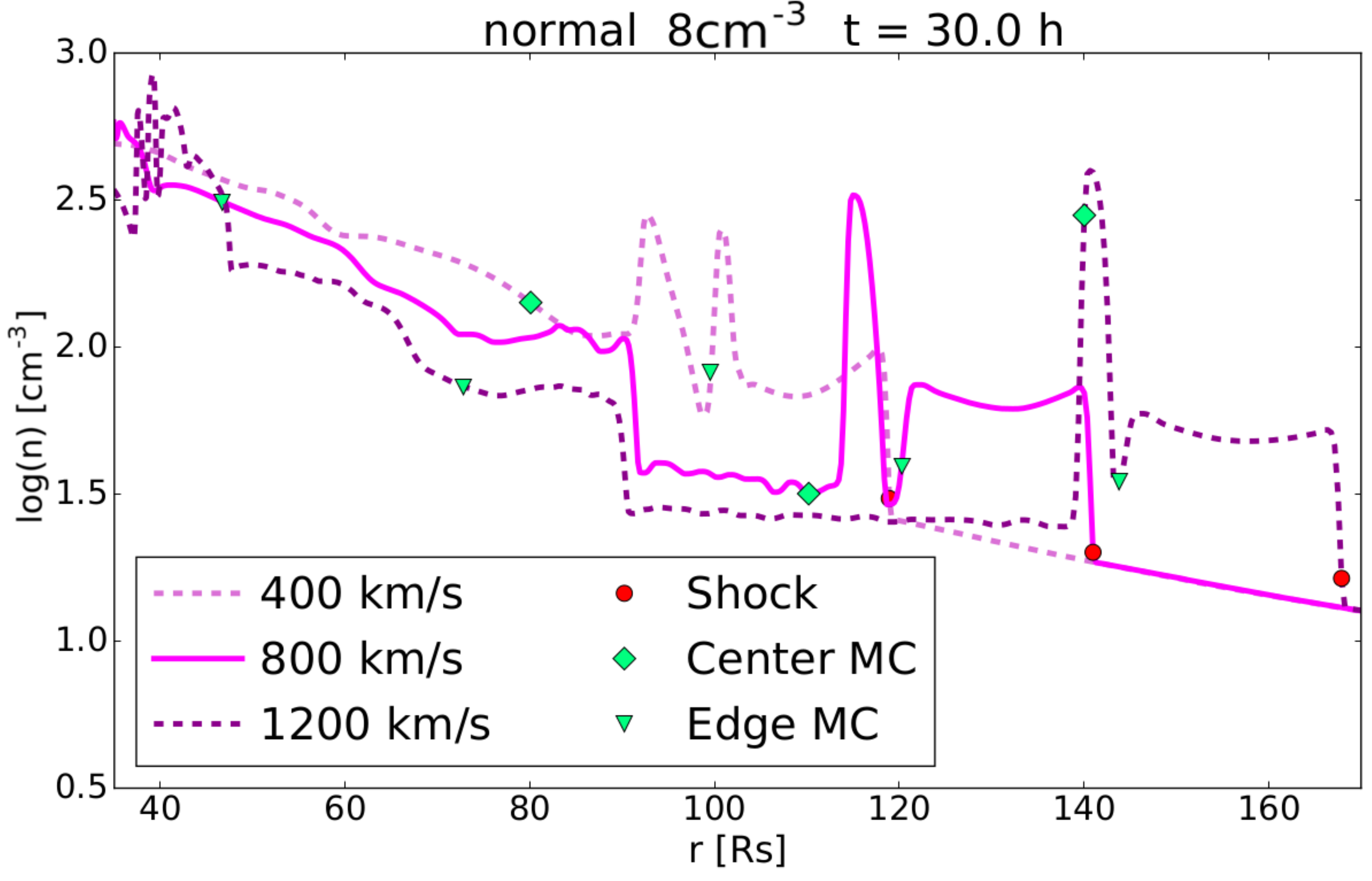}
\includegraphics[width=0.495\linewidth]{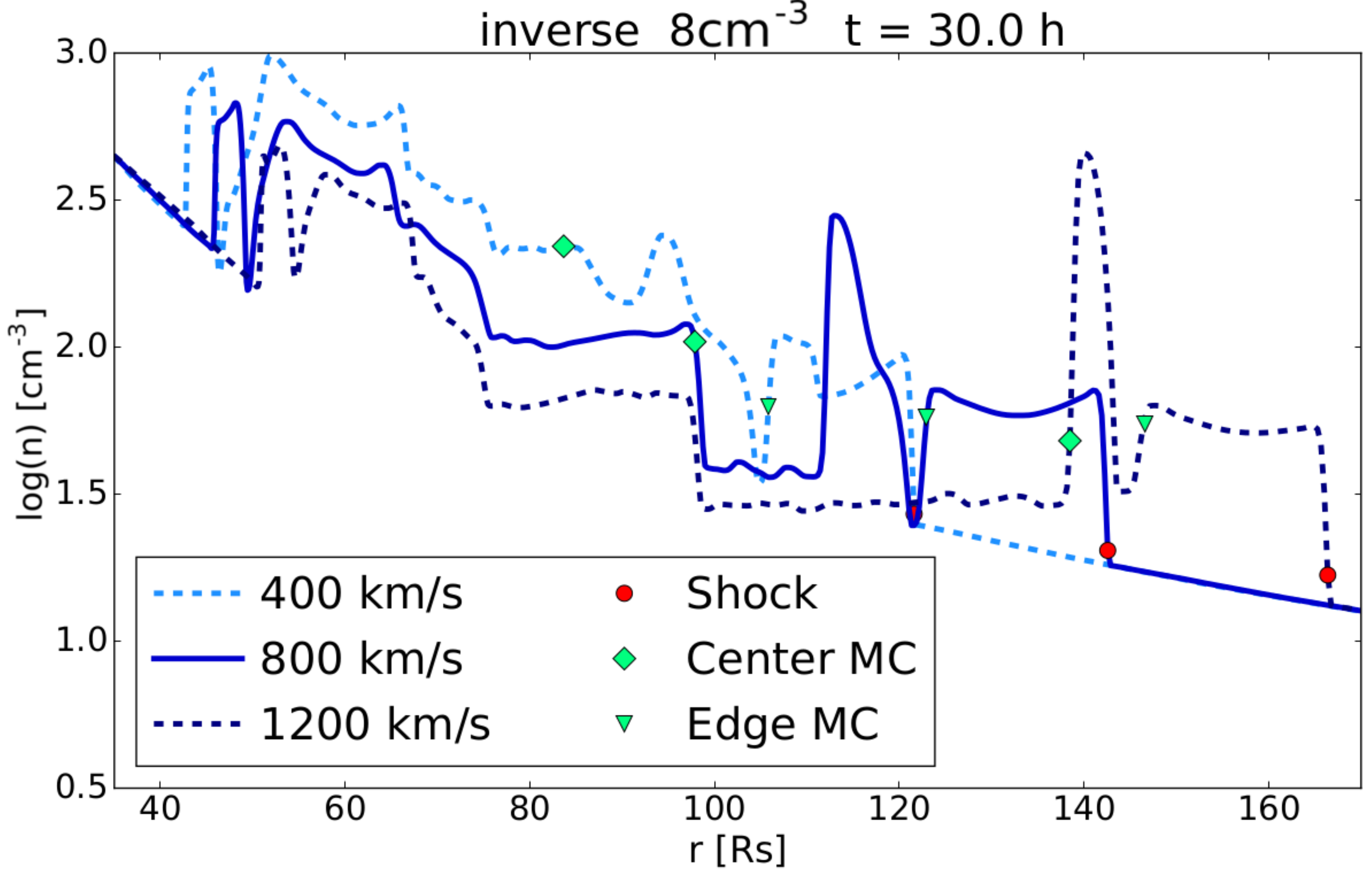}\par
\includegraphics[width=0.495\linewidth]{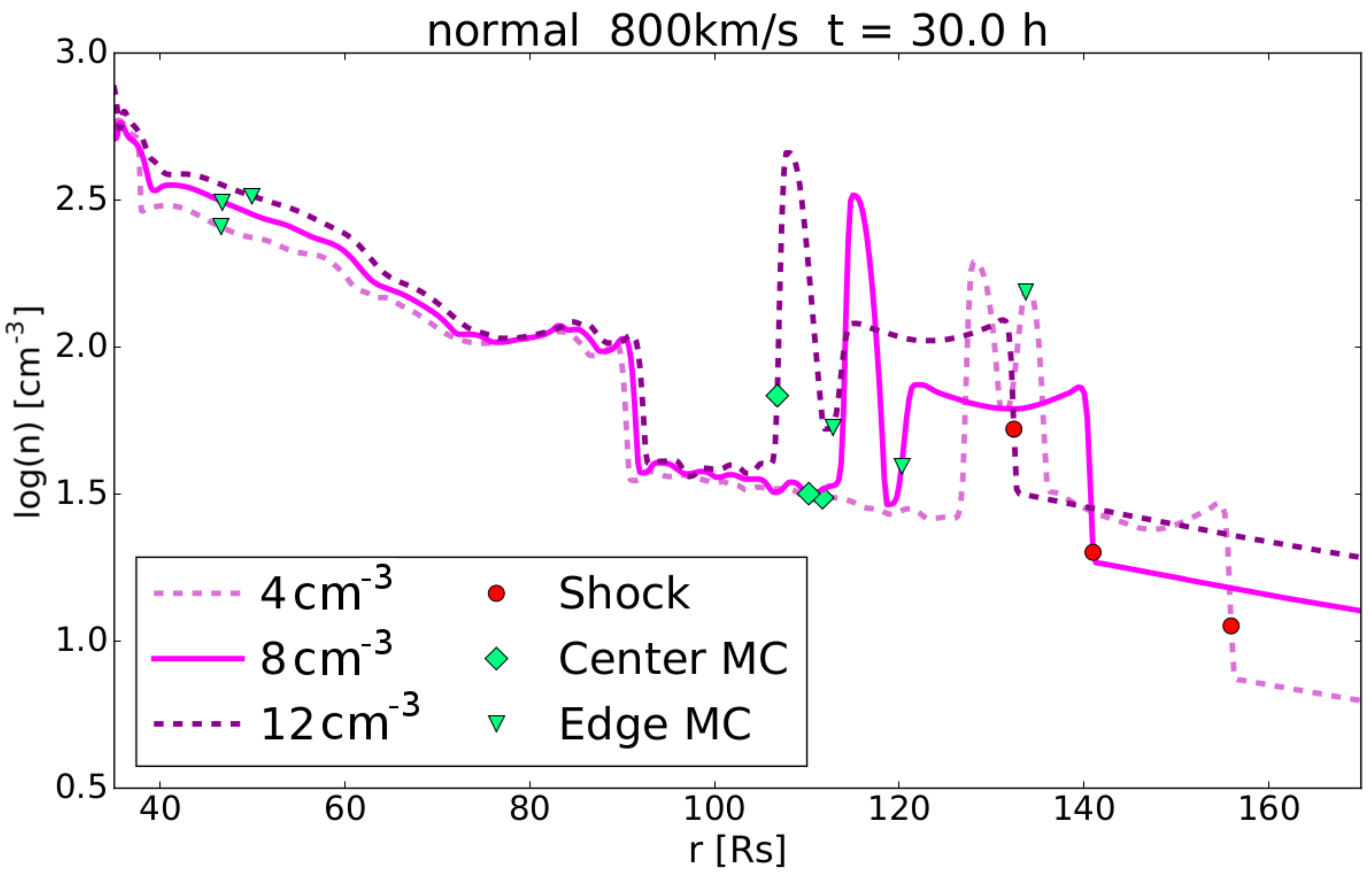}
\includegraphics[width=0.495\linewidth]{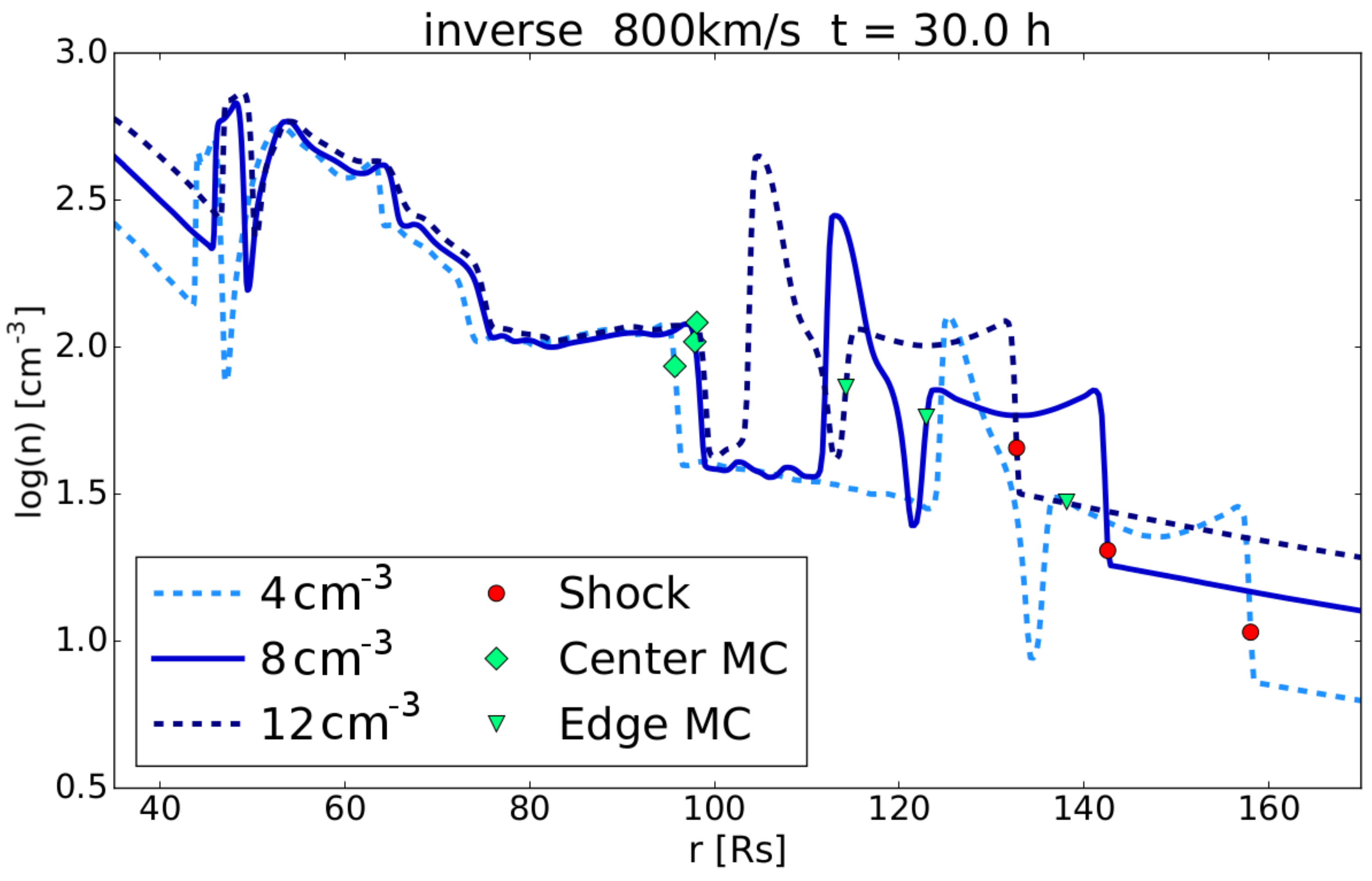}}
\caption{Number density of the radial cross section at the equator $30\;$h after ejection for normal CMEs in the left column and inverse CMEs in the right column. In the upper panels a medium density background wind was chosen for three different initial velocities of the CME. In the lower panels, a CME with an initial velocity of $800\;$km/s was ejected in three different background winds (indicated with the different line styles). The red dots represent the location of the CME shock front, the green diamonds represent the position of the centre of the magnetic cloud, and the green triangles represent the edges of the magnetic cloud.}\label{fig:cross_section}
\end{figure*}

\begin{figure}[!htb]
\begin{center}
\includegraphics[width=0.47\linewidth]{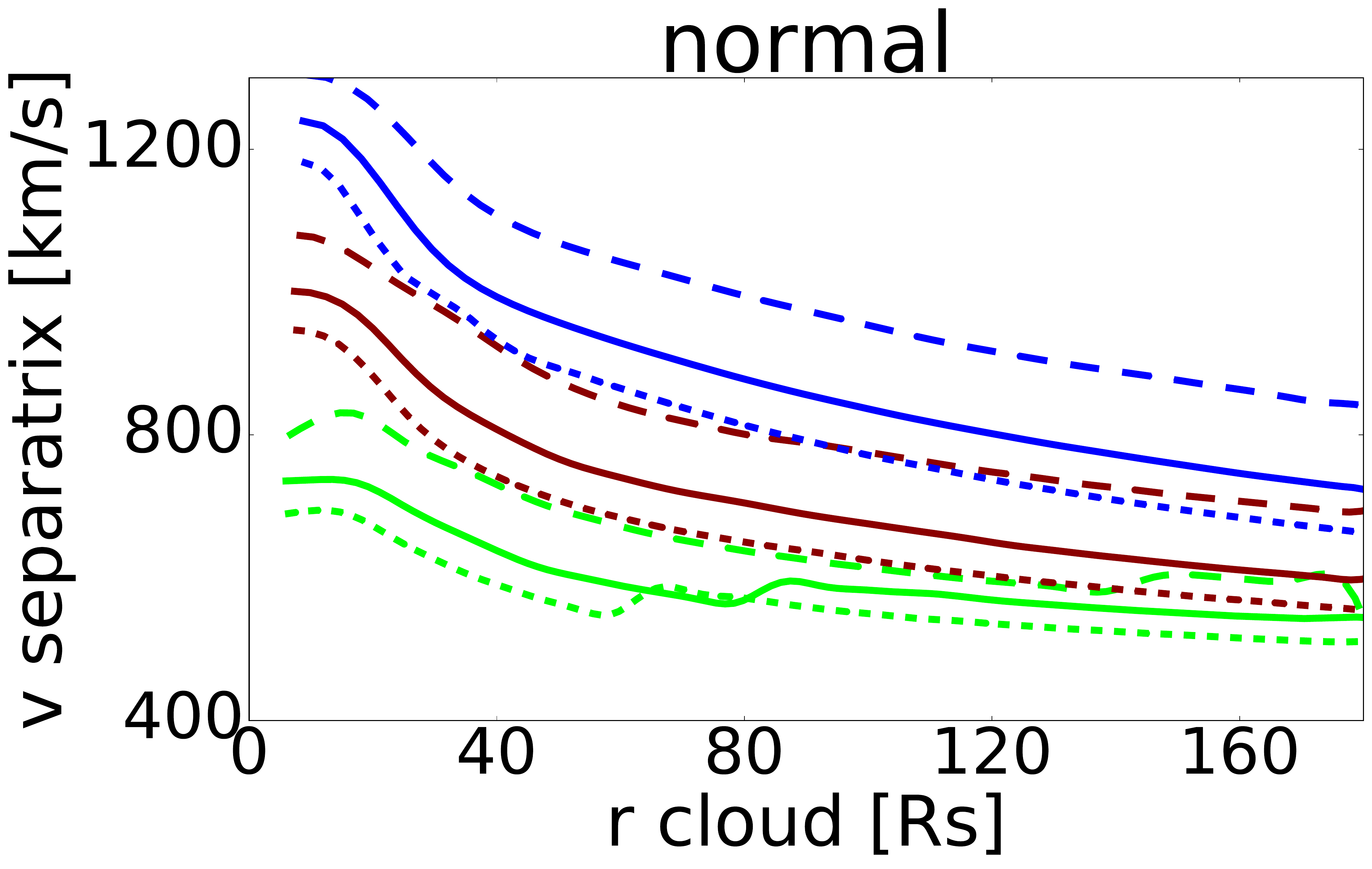}
\includegraphics[width=0.47\linewidth]{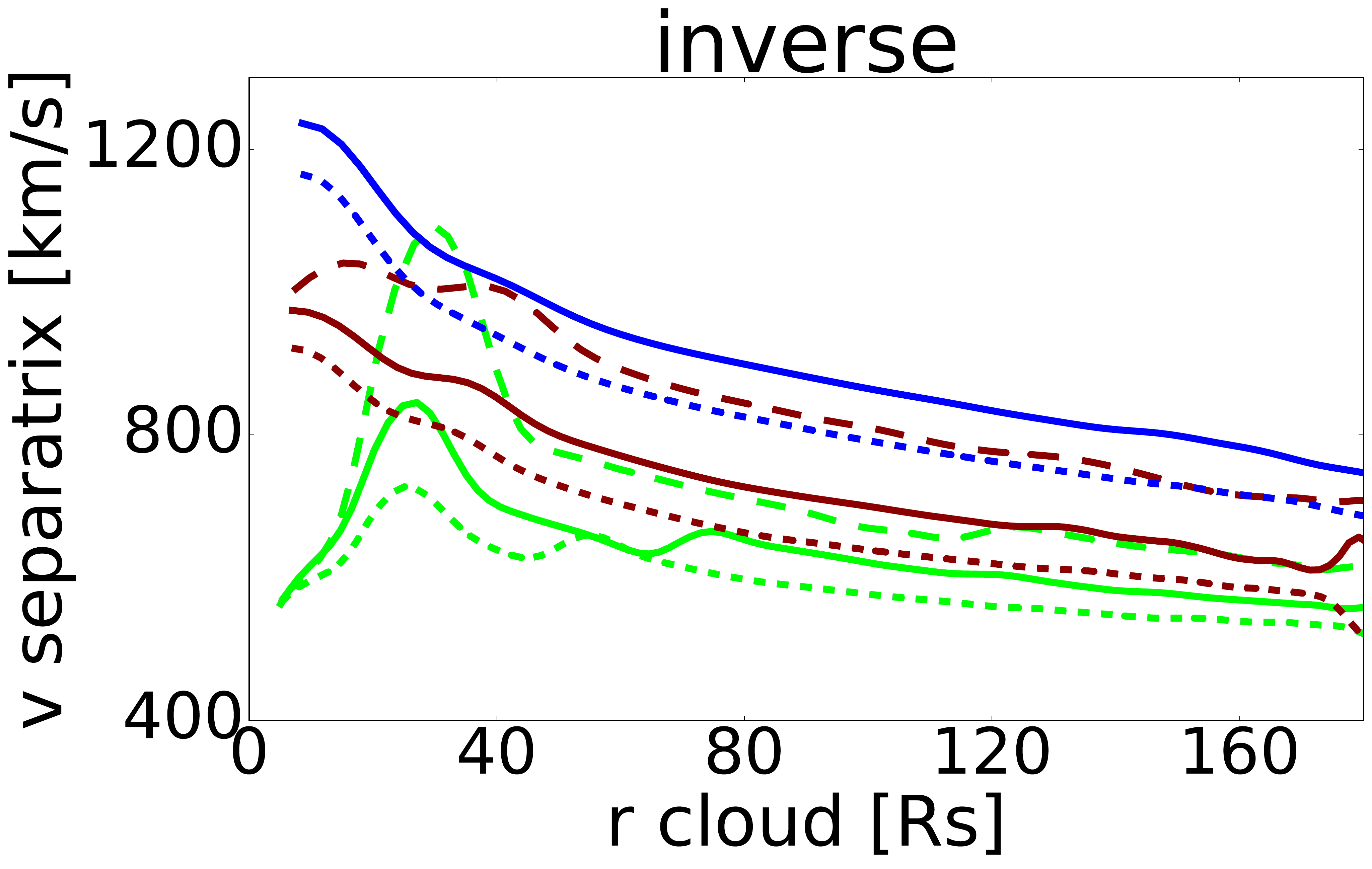}
\includegraphics[width=0.47\linewidth]{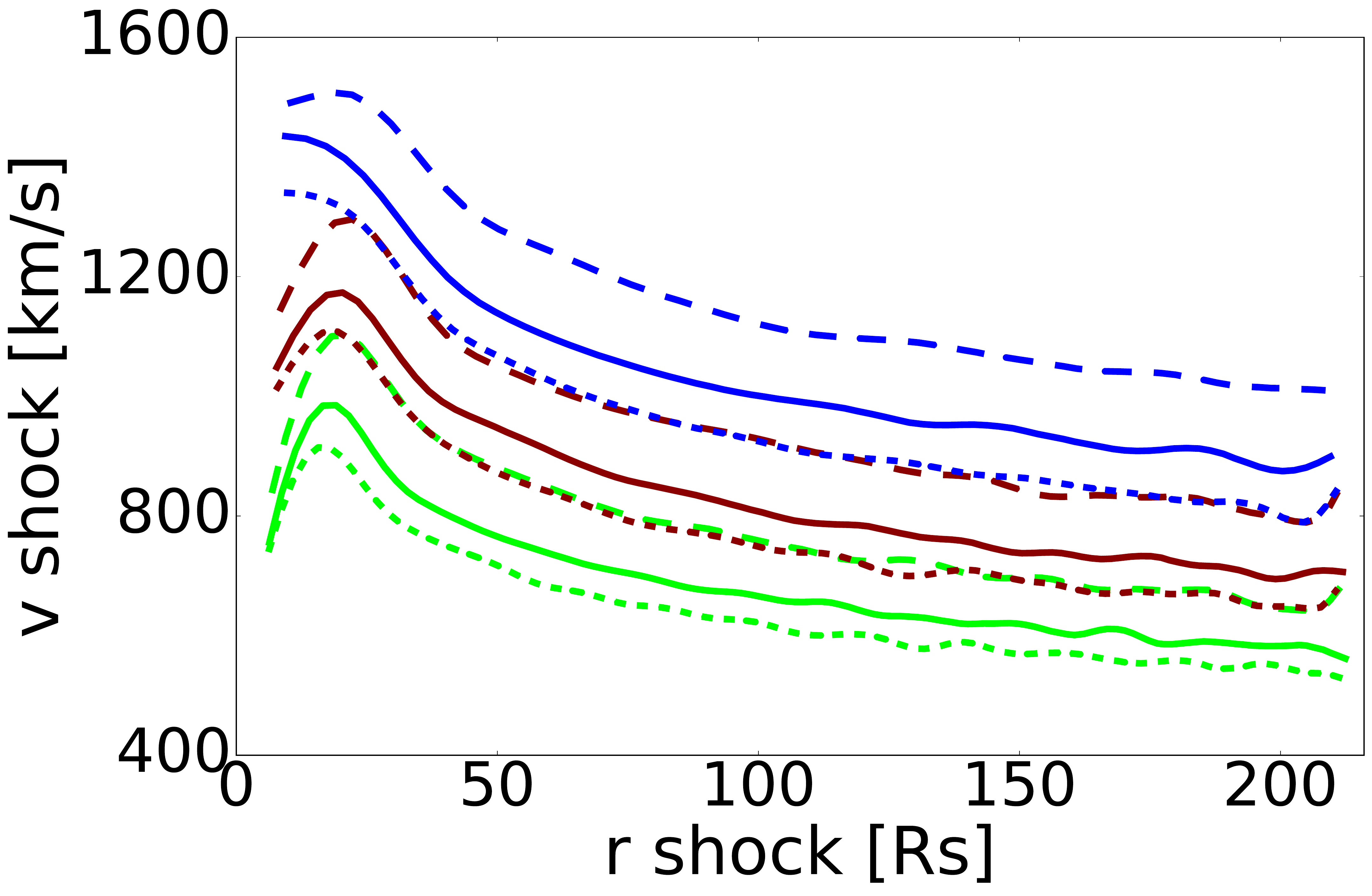}
\includegraphics[width=0.47\linewidth]{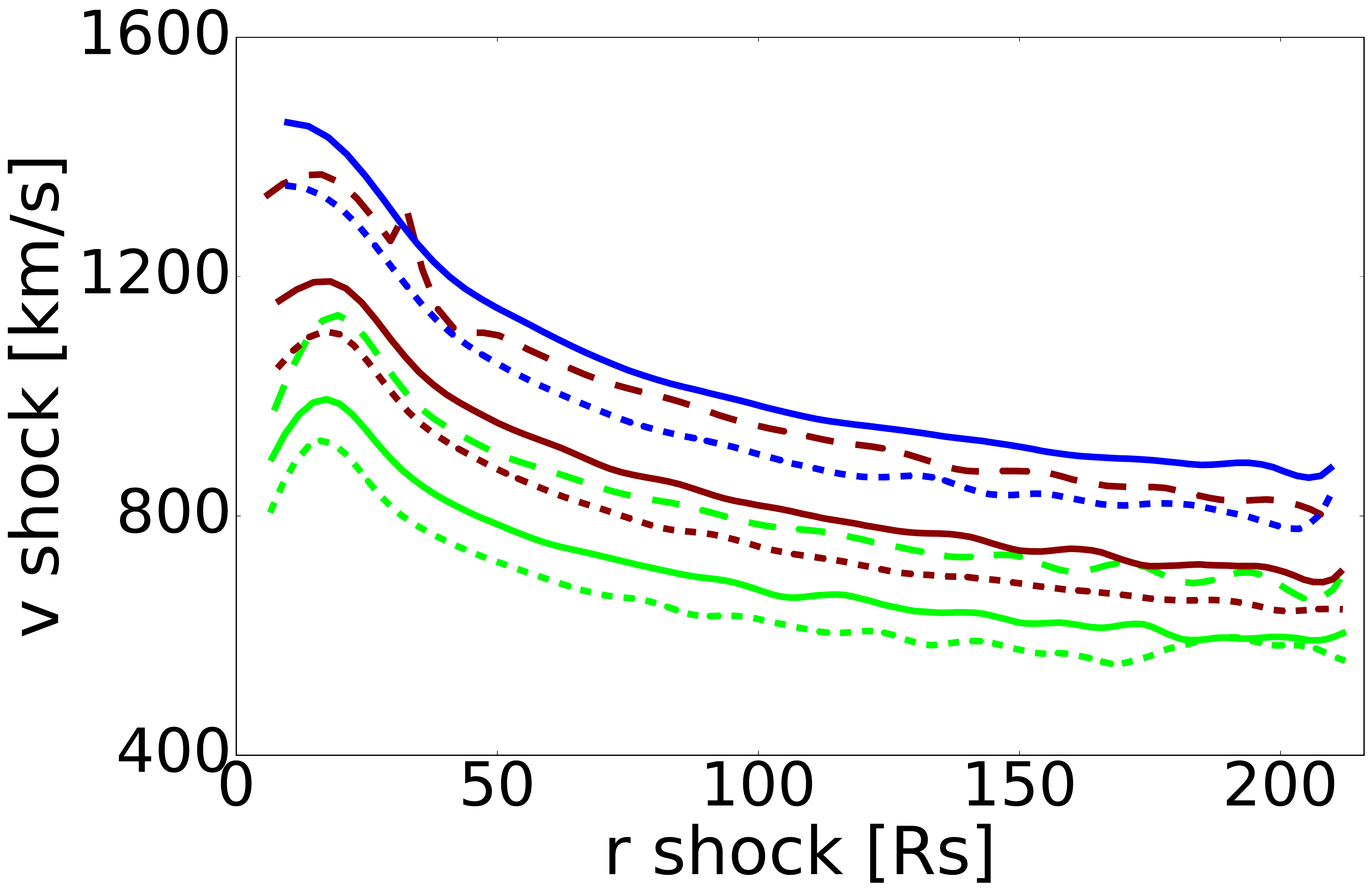}
\caption{Velocity profiles of both the front of the separatrix (upper panels) and the shock (lower panels) in function of their position. The left panels contain the results of all normal CME simulations while the right panels contain the results of the inverse CME simulations. The colours represent different initial velocities (green: $400\;$km/s, red: $800\;$km/s, blue: $1200\;$km/s), while the line styles represent different background wind densities (dashed: $4\;$cm\textsuperscript{-3}, solid: $8\;$cm\textsuperscript{-3}, dotted: $12\;$cm\textsuperscript{-3}. Both columns have the same scale to facilitate comparisons.} \label{fig:vel_profiles}
\end{center}
\end{figure}

\begin{table}
\centering
\begin{tabular}{|c|c|c|c|}
\hline 
\textbf{NORMAL}  & 400 km/s & 800 km/s & 1200 km/s \\
\hline 
4 cm\textsuperscript{-3} & 53 & 43.5 & 35 \\
\hline 
8 cm\textsuperscript{-3} & 60.5 & 49.5 & 39.5 \\
\hline 
12 cm\textsuperscript{-3} & 65 & 53 & 43 \\
\hline 
\end{tabular}  
\par
\vspace{0.2cm}
\centering
\begin{tabular}{|c|c|c|c|}
\hline 
\textbf{INVERSE}  & 400 km/s & 800 km/s & 1200 km/s \\
\hline 
4 cm\textsuperscript{-3} & 51 & 43 & - \\
\hline 
8 cm\textsuperscript{-3} & 59 & 49 & 40 \\
\hline 
12 cm\textsuperscript{-3} & 63.5 & 53.5 & 44 \\
\hline 
\end{tabular} \caption{Shock arrival times at 1~AU expressed in hours after ejection for both normal (upper table) and inverse (lower table) CMEs.}\label{table:arrival_shock}
\end{table}

\begin{table}
\centering
\begin{tabular}{|c|c|c|c|}
\hline 
\textbf{NORMAL}   & 400 km/s & 800 km/s & 1200 km/s \\
\hline 
4 cm\textsuperscript{-3} & 64.5 & 52.5 & 42.5 \\
\hline 
8 cm\textsuperscript{-3} & 70 & 59.5 & 48.5 \\ 
\hline 
12 cm\textsuperscript{-3} & 73.5 & 64 & 52 \\
\hline 
\end{tabular}  
\par
\vspace{0.2cm}
\centering
\begin{tabular}{|c|c|c|c|}
\hline 
\textbf{INVERSE}  & 400 km/s & 800 km/s & 1200 km/s \\
\hline 
4 cm\textsuperscript{-3} & 60 & 51 & - \\
\hline 
8 cm\textsuperscript{-3} & 67 & 58.5 & 47.5 \\
\hline 
12 cm\textsuperscript{-3} & 72 & 63 & 51.5 \\
\hline 
\end{tabular} \caption{Magnetic cloud arrival times at 1~AU expressed in hours after ejection for both normal (upper table) and inverse (lower table) CMEs.}\label{table:arrival_cloud}
\end{table}

Figure~\ref{fig:cross_section} shows the variation in the number density along the radial cross section at the equator, $30\;$h after the ejection. The left panel shows this for normal CMEs and the right panel for inverse CMEs. The upper figures result from CMEs ejected with different initial velocities in a medium density wind while the bottom row shows results of CMEs ejected with an initial velocity of $800\;$km/s in varying background winds. The locations of the shock front and the centre of the magnetic cloud are shown in each simulation, as are the left and right boundaries of the magnetic cloud. From the upper panel we can see that, in addition to yielding an earlier arrival, the centre of the magnetic cloud of CMEs with a higher initial velocity are located more towards the front of the separatrix. This results in higher density peaks between the separatrix and the centre of the magnetic cloud and the initially circular form of the separatrix becomes more flattened as a larger difference between the magnetic cloud speed and the solar wind leads to a higher drag force acting upon the magnetic cloud;  the drag force is proportional to the square of the relative velocity (i.e. the difference between the CME velocity and the velocity of the background wind). The lower panel shows that a higher density background wind leads to lower arrival times (due to a higher drag force, which is also proportional to the density of the background wind), and the separatrix thus becomes more compressed. A different background wind density will affect the position of the front of the CME, while the back edge position is approximately the same for the three simulations. However, a different initial velocity leads to a compression or inflation over the whole volume of the magnetic cloud. The evolution of the density inside the magnetic cloud, initially the same for all CMEs, thus depends  on the background wind and on the initial speed. The density peak becomes higher for a higher background wind density and/or a higher initial velocity. Comparing normal and inverse ICMEs, we see that the shock, sheath, and front part of the magnetic cloud are quite similar in density structure, but the tail region of the magnetic cloud is more turbulent and higher in density for inverse ICMEs due to the magnetic reconnection occurring in the rear. \par

Knowing the position of the shock and of the front and centre of the magnetic cloud at all times allows the determination of their velocity. Despite the four levels of mesh refinement, there is some scatter in the velocity determination, which becomes  larger moving away from the Sun because the grid cells  increase in size with radial distance from the Sun. This scatter is reduced by employing a Savitsky-Golay filter to smooth the data. \par

\begin{figure}[!htb]
  \begin{center}
    \includegraphics[width=\hsize]{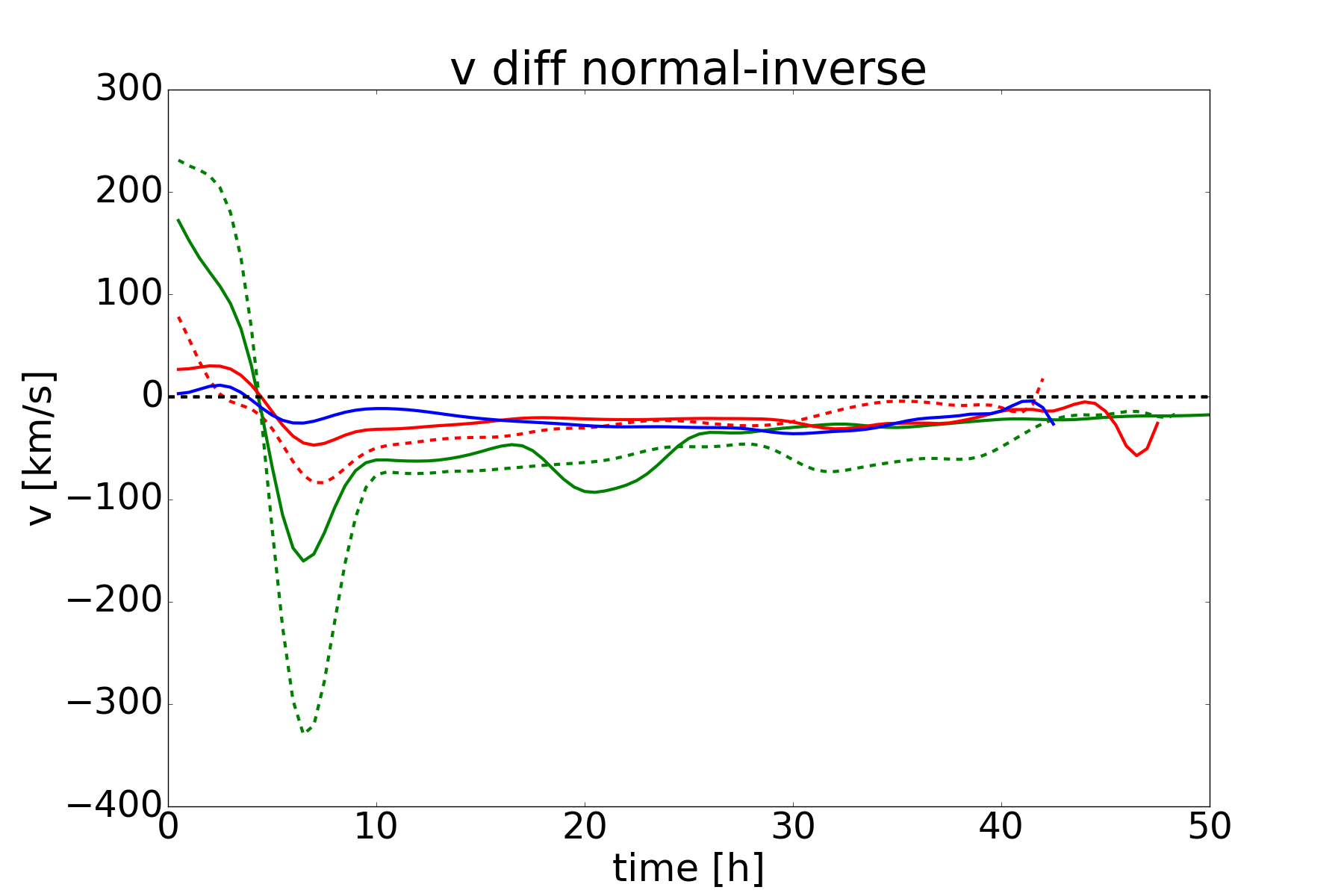}
  \end{center}
  \caption{Velocity difference of the MC front between normal and inverse CMEs with the same initial velocity ejected in the same background wind vs. time.   The colours represent different initial velocities (green: $400\;$km/s, red: $800\;$km/s, blue: $1200\;$km/s), while the line styles represent different background wind densities (dashed: $4\;$cm\textsuperscript{-3}, solid: $8\;$cm\textsuperscript{-3}). The black dotted line represents v\textsubscript{diff}=0.}\label{fig:vdiff}
\end{figure}

Figure~\ref{fig:vel_profiles} shows the velocity of the front of the magnetic cloud versus its position in the upper panel and the velocity of the shock versus its position in the lower panel. It can be seen that the speed of the separatrix almost immediately decelerates, except for the case of the inverse ICMEs in a low density solar wind, which will be discussed below. We note that the magnetic clouds of inverse ICMEs seem to have a higher speed in all cases once the CME has passed 30~R$_\odot$. The order of the different velocity curves stays the same for the whole duration, meaning that ICMEs  with a higher initial velocity keep a higher velocity throughout their evolution and higher background wind densities result in a lower ICME speed throughout their entire propagation. We  also note that the shock fronts experience a strong acceleration in the beginning of its propagation. This acceleration is stronger for lower Alfv\'en Mach numbers (i.e.\ lower surrounding wind densities and/or lower initial CME velocities). Tables~\ref{table:arrival_shock} and ~\ref{table:arrival_cloud} show the arrival time at 1~AU of the shock and of the front of the magnetic cloud, respectively. Inverse CMEs have a higher separatrix velocity compared to normal CMEs, which is surprising considering that previous studies have claimed the opposite \citep{Chane2006}, explaining the magnetic field lines being curved behind a normal CME create a magnetic slingshot effect. In our simulations, the speeds of the separatrix of normal CMEs are initially higher, but after approximately 20-25 R$_\odot$ inverse separatrix speeds become higher than their normal counterparts, as can be seen in Figure \ref{fig:vdiff}. Positive values mean the normal MC fronts have a higher velocity than inverse MCs. We note that the difference is taken at the same time after ejection so the MC fronts are not at exactly the same position. The figure shows the competing effects of two different phenomena: (1) the slingshot effect that makes normal CMEs travel faster, and (2) magnetic reconnection (in the tail for inverse and in the front for normal CMEs) increases the size of the magnetic cloud of inverse CMEs or decreases it for normal CMEs. Magnetic reconnection in the rear of inverse CMEs adds field lines to the magnetic cloud, propagating the front of the separatrix forward, even in the frame of the CME. On the other hand, normal CMEs are   stripped of magnetic field due to magnetic reconnection occurring at their front, diminishing the size of the magnetic cloud and, in the frame of the CME, receding the front of the separatrix. \par

\subsection{Deformation}\label{subsec:deformation}

\begin{figure}[!htb]
\begin{center}
\includegraphics[width=0.47\linewidth]{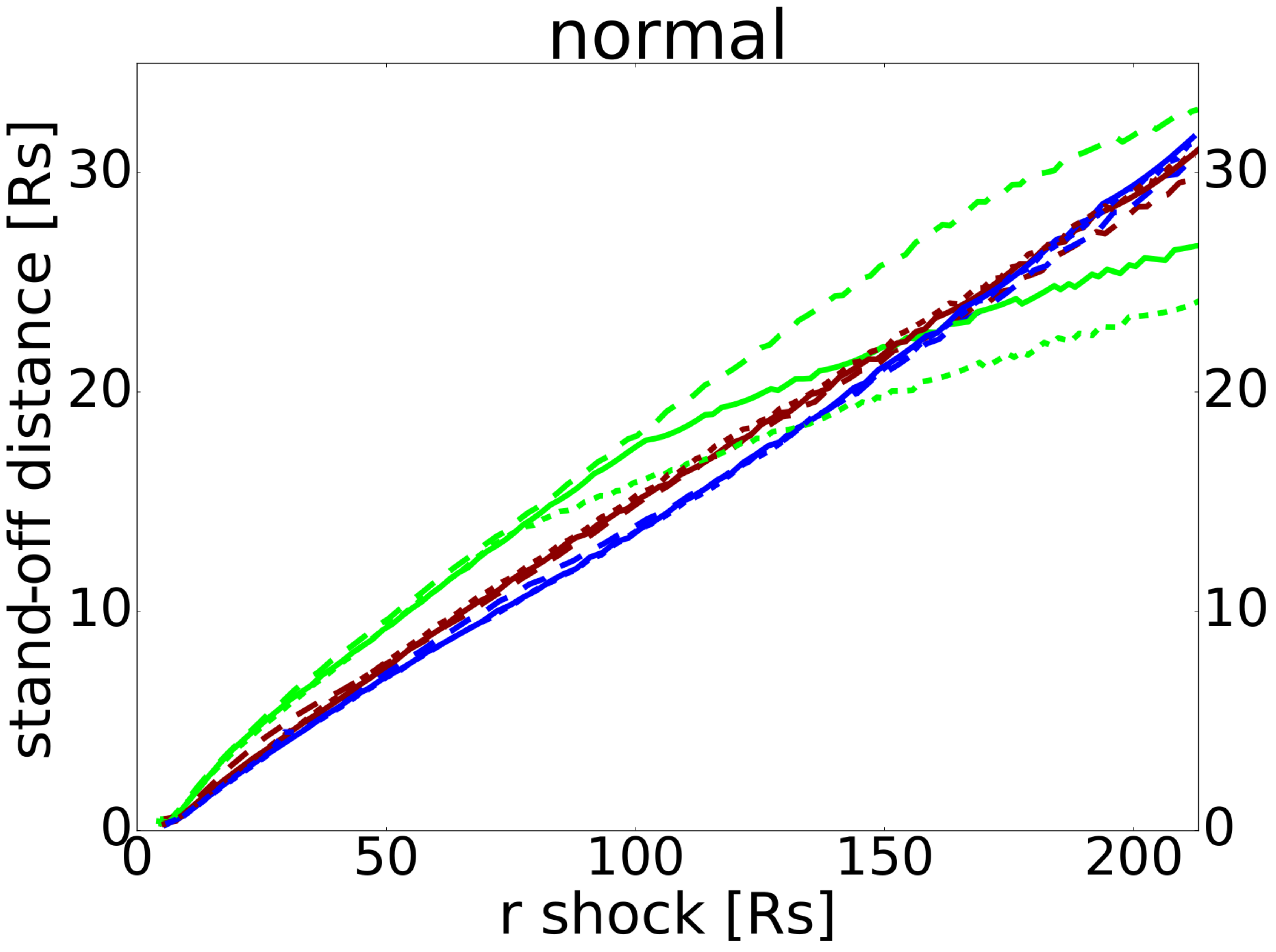}
\includegraphics[width=0.47\linewidth]{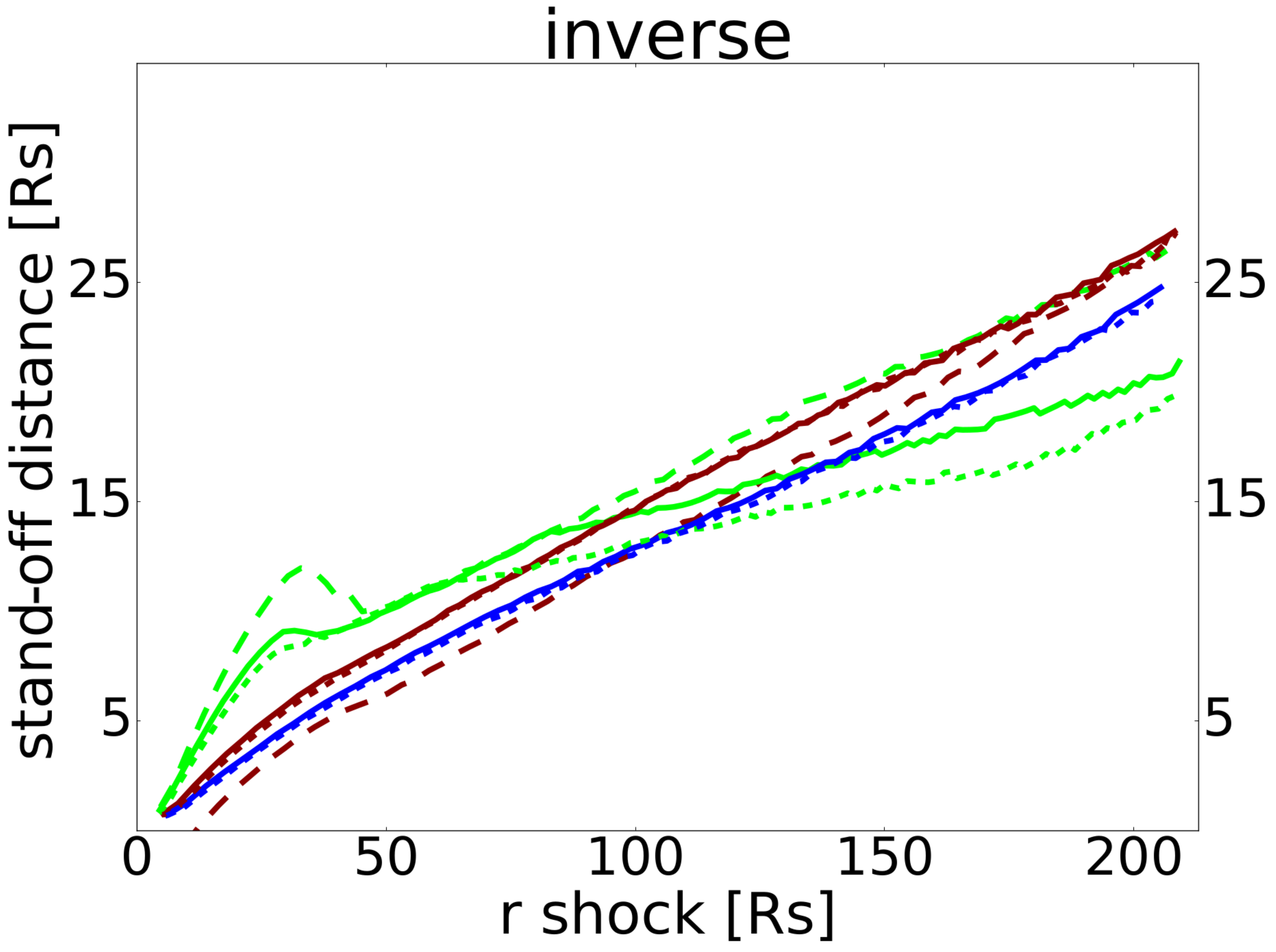}
\includegraphics[width=0.47\linewidth]{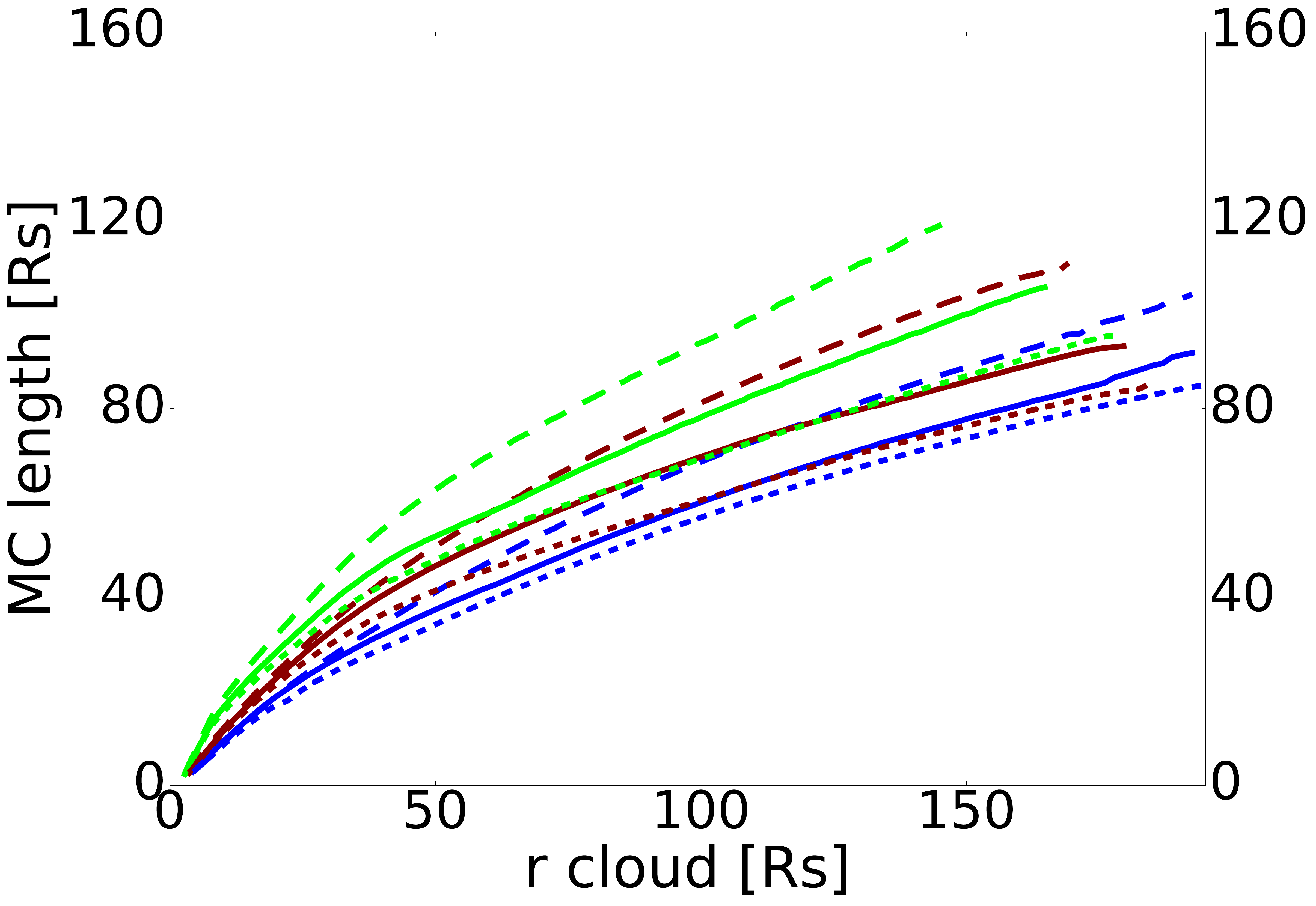}
\includegraphics[width=0.47\linewidth]{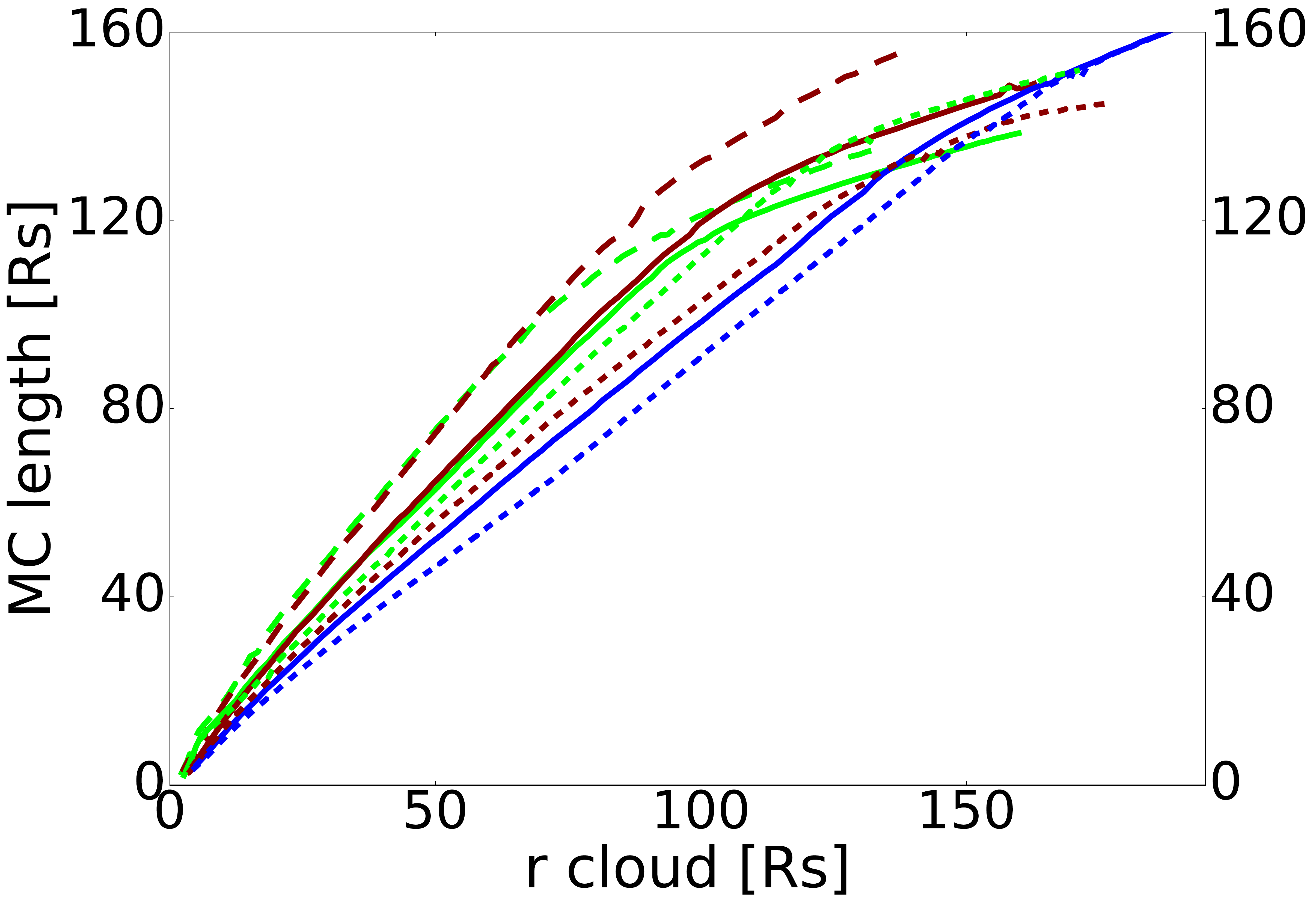}
\includegraphics[width=0.47\linewidth]{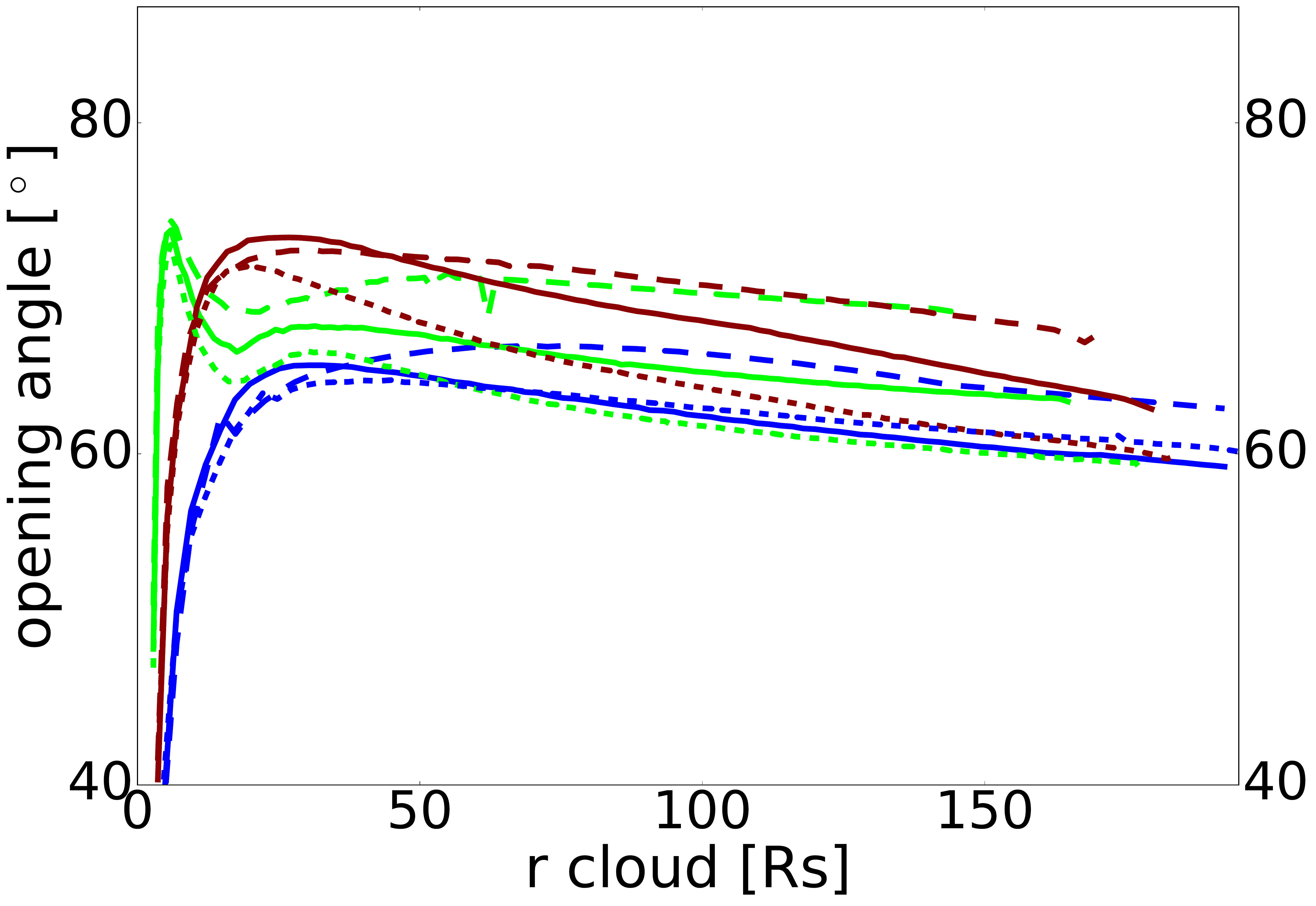}
\includegraphics[width=0.47\linewidth]{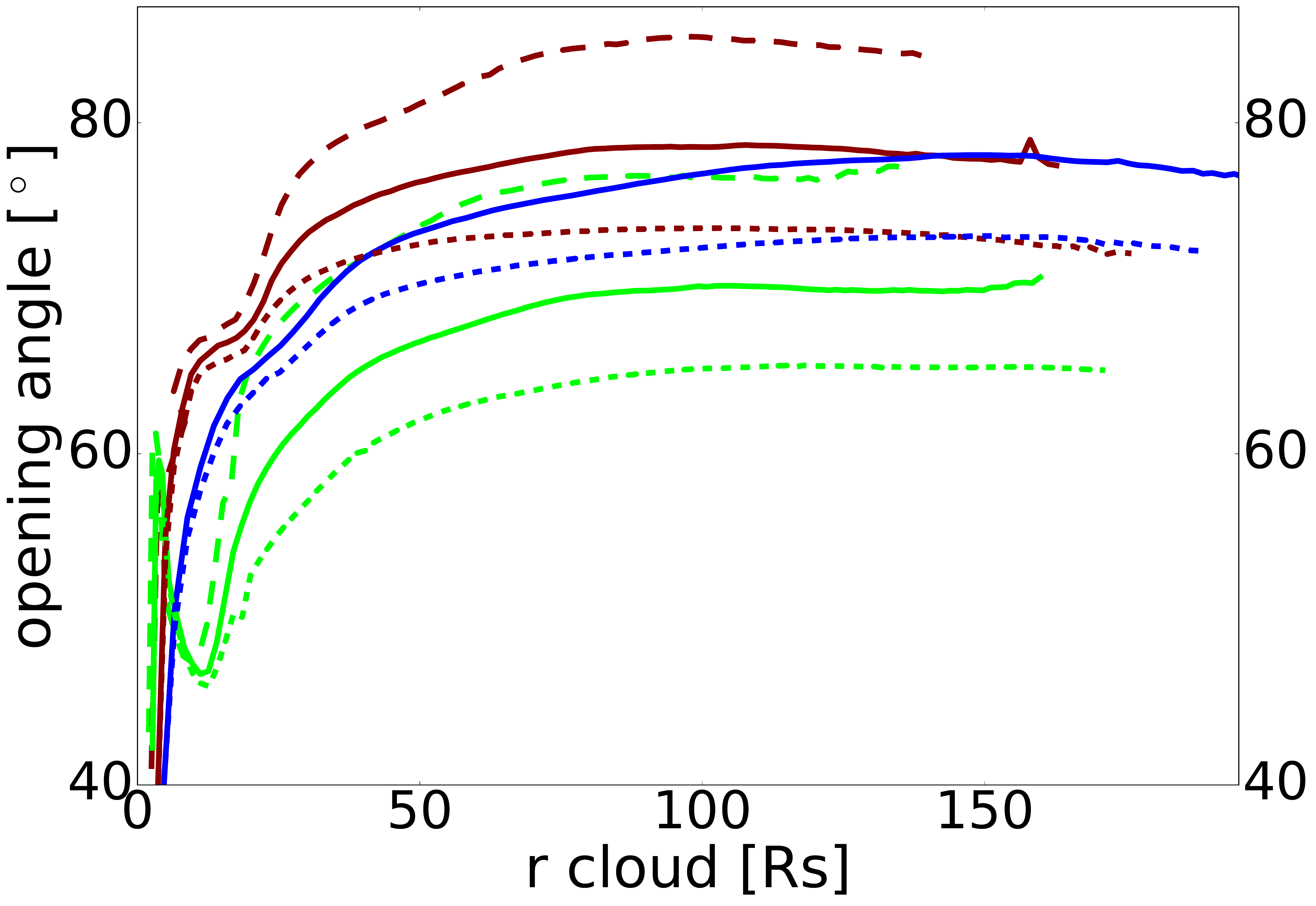}
\caption{   Properties for all normal CMEs  and for all inverse CMEs (left and right, respectively). The colours represent different initial velocities (green: $400\;$km/s, red: $800\;$km/s, blue: $1200\;$km/s), while the line-styles represent different background wind densities (dashed: $4\;$cm\textsuperscript{-3}, solid: $8\;$cm\textsuperscript{-3}, dotted: $12\;$cm\textsuperscript{-3}). From top to bottom, the $y$-axes represents the stand-off distance, the length of and the height of the separatrix of the CME, respectively. The $x$-axis in the first row represents the position of the shock, while the $x$-axes of the second and third rows represent the position of the centre of the magnetic cloud. Both columns have the same scale to facilite comparisons.} \label{fig:standoff}
\end{center}
\end{figure}

The upper row of Figure~\ref{fig:standoff} shows the evolution for the stand-off distance (i.e.\ the width of the sheath), for all normal and inverse CME simulations. It is immediately clear from this figure that the initial speed of the ICME and solar wind density have little influence on the stand-off distance, except for slow CMEs, where the effect of the drag is smaller compared to the other simulations. On the other hand, the polarity of the eruption seems to have a large effect on the stand-off distance, with inverse CMEs having a considerable lower stand-off distance than their normal counterparts. Due to magnetic reconnection occurring at the front, normal CMEs experience erosion of their magnetic clouds, stripping away magnetic field lines and thus increasing the stand-off distance. Inverse CMEs do not suffer from this effect. In fact, magnetic reconnection happening at the tail of inverse CMEs may add magnetic field lines to the magnetic cloud.
The middle and lower rows of Figure~\ref{fig:standoff} show the length and the opening angle of the magnetic clouds, respectively, for normal (left) and inverse (right) CMEs. The left panel of the middle row shows that normal CMEs ejected in the same background wind with a higher initial velocity have a smaller length when their magnetic cloud reaches the same distance from the Sun. This is due to the fact that CMEs with lower velocities have more time to expand. Similarly, CMEs with the same velocity in a higher density wind experience a stronger drag than those ejected in lower density winds. This causes longer travel times, but at the same time the magnetic cloud also becomes more compressed. We see a similar trend for the $800\;$km/s and $1200\;$km/s inverse CMEs in the right panel, but the low velocity CMEs (green curves) seem to behave differently, which will be discussed below. Comparing the left and right panels of the figure, it is clear that inverse CMEs are much more elongated in the radial direction. Normal CMEs fall in a range of approximately $80-120\;R_{\odot}$ at 1~AU, while inverse CMEs reach radial lengths between approximately $130-160\;R_{\odot}$. Even for normal CMEs, this is about twice the typical CME length (\textasciitilde40-50~R$_\odot$). This can be attributed to our very high initial CME density and pressure leading to an extreme expansion. It must be noted that the radial length of the CME is measured along the equator, where the CME cross section is the longest. The extreme length difference is mainly due to the relatively extended tail end of inverse CMEs and the difference becomes smaller further away from the equator, as can be seen in Figure~\ref{fig:together}. \par

The lower row of Figure~\ref{fig:standoff} shows the evolution of the opening angle of the CMEs, which is here defined as the latitudinal width of the separatrix. Similarly to the effect on the CME radial length, lowering the background wind density for both normal and inverse CMEs with the same initial speed results in larger opening angles.
Normal CMEs initially expand faster than inverse CMEs, but beyond approximately 20~R$_{\odot}$ normal CMEs experience a decreasing opening angle while inverse CMEs keep expanding until  50~R$_{\odot}$, beyond which their opening angle becomes constant. We note that in both the normal and in the inverse CME figure the green curves (indicating the low initial CME velocity) display a different behaviour at the beginning of their propagation (below 20~R$_{\odot}$) compared to the red and blue curves (indicating medium and high CME velocities, respectively). At 1~AU, the opening angle of the normal CMEs in our simulations falls between approximately 60$\degree$ and 70$\degree$, and the spread angles of inverse CMEs range from approximately 65$\degree$ to 85$\degree$. We must keep in mind, however, that this refers to the opening angle of the separatrix, and that the width of the full ejection for normal CMEs is actually larger due to the magnetic structures above and below the separatrix. The top row of Figure~\ref{fig:together} shows that normal CMEs develop small magnetic structures at the sides of the magnetic cloud, whose pressure restrains the angular expansion of the magnetic cloud. 
  \par

\begin{figure}[!htb]
    \centering
    \includegraphics[width=\hsize]{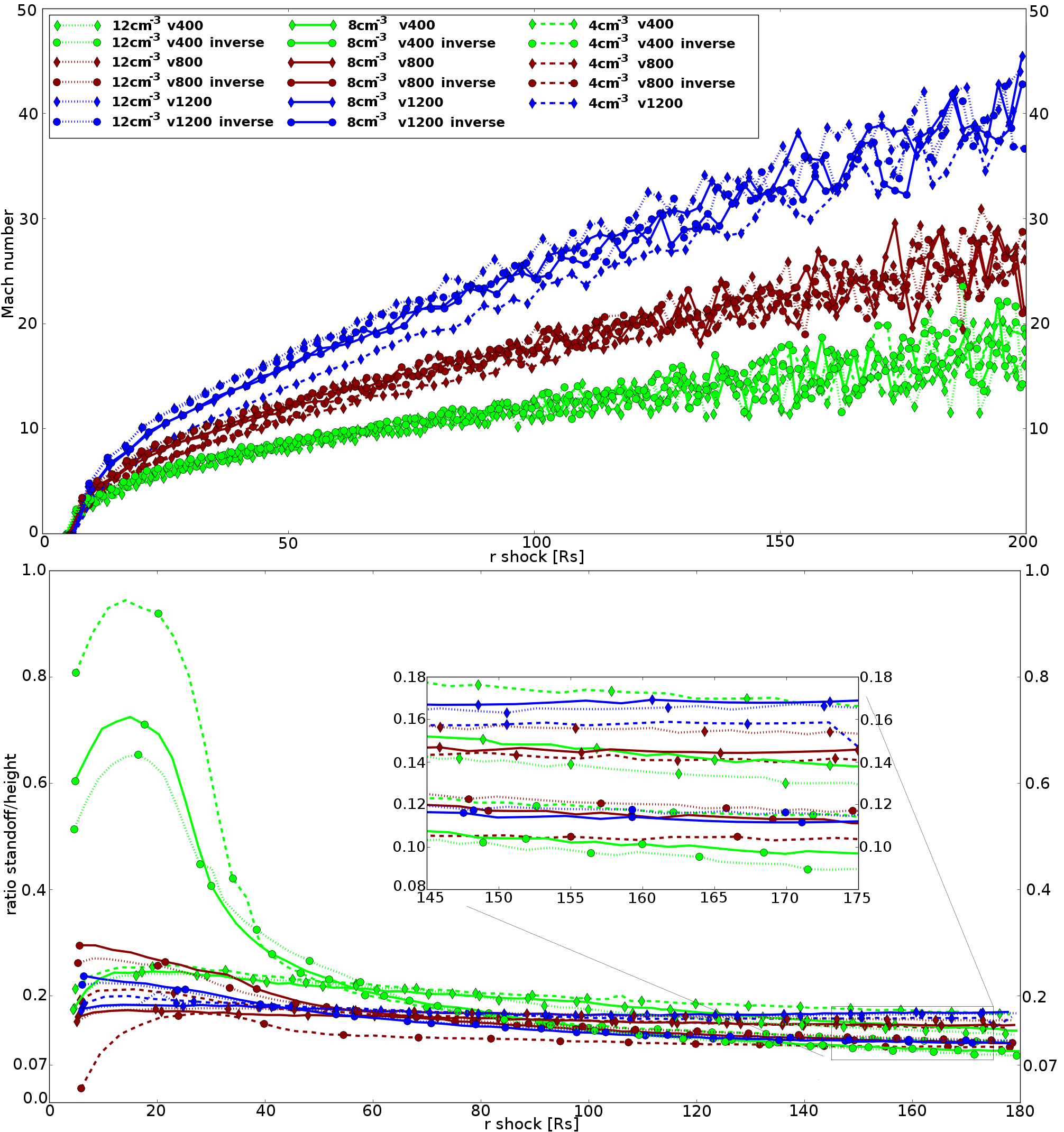}
    \caption{Upper panel: Alfv\'en Mach number at the location of the shock along the ecliptic for all simulations. Lower panel: Ratio stand-off distance to height. Polarity is represented by  diamonds  for normal CMEs and dots for inverse CMEs (one symbol is shown for every five data points).}
    \label{fig:Mach}
\end{figure}

Coronal mass ejection  stand-off distances are often investigated in relation to the Mach number or the radius of curvature of the CME. The upper panel of Figure \ref{fig:Mach} shows the Mach number 7.25 degrees above the equatorial plane, which is the obliquity of the Sun ($\frac{v_{shock}-v_{SW}}{Va_{SW}}$). This direction was preferred to that along the equator due to the very low $B$-magnitude in the current sheet, which leads to extreme values for the Alv\'en speed and consequently for the Mach number as well. The figure shows that the Alfv\'en Mach numbers of the simulations can be divided in three distinct groups based on the initial CME speed. Both polarity and background wind density appear to affect the Mach number to a much lesser degree. It may seem surprising that the background wind density does not noticeably influence the Mach number, even though each background wind has a different Alfv\'en speed. However,  while a higher wind density implies a lower Alv\'en speed, the velocity difference between the shock and the solar wind is lower as well due to a higher drag force acting upon the CME. The increasing scatter of the curves as the shock moves further away from the Sun is due to increasing grid cell sizes for larger radial distances. These Mach number values (\textasciitilde15-45 at 1~AU) are considerably higher than observed values (typically \textasciitilde4-6 at 1~AU). This is due to our background solar wind simulations having a lower Alfv\'en speed at 1~AU ($\pm$9~km/s, $\pm$11~km/s, and $\pm$14~km/s for our high, middle, and low density background winds, respectively) than typically measured, which is approximately 40~km/s. Due to the symmetry in our idealised set-up the Earth is always close to the current sheet,  hence the low magnetic field strength and Alfv\'en velocity. Our CMEs also experience less deceleration than typical CMEs, possibly due to their very high initial momentum, making them harder to slow down. An investigation was performed by \cite{Siscoe2008} where the stand-off distance of the CME was normalised to the radius of curvature of the leading edge. They distinguished between two types of sheath regions, namely the `expansion sheath' and the `propagation sheath'. The former is a sheath region around an object that expands into a solar wind, but does not propagate through it, causing the  solar wind to pile up in front of the sheath. The latter refers to a sheath region of an object that moves through the solar wind, where the solar wind enters the sheath region and  flows around the object. Since the CME sheaths both propagate though the solar wind and expand  at a non-negligible rate compared to its propagation speed, they state that the CME sheaths contain properties of both propagation and expansion sheaths. They conclude for high Mach numbers (M $>$ 5, which is clearly satisfied, as shown in the upper panel) that the stand-off distance normalised to the radius of curvature should fall between approximately 0.07 (for expansion sheaths) and 0.2 (for propagation sheaths). In a similar manner to an investigation into stand-off distances for 2.5D CME simulations done by \cite{Savani2012}, we   take the vertical height of the separatrix as a proxy for the radius of curvature. As seen in Figure \ref{fig:standoff}, the slow inverse CMEs have a much lower opening angle (which can be seen as a measure for the CME height) in the beginning of their propagation than the other simulations. This leads to extremely high ratios of stand-off distance to CME height, even almost reaching one in the case if the low density background wind. Most of the simulations start with a ratio between 0.18 and 0.30 which then relaxes to values between 0.09 and 0.18, in agreement with \cite{Siscoe2008}. The zoom inset in the lower panel of Figure \ref{fig:Mach} shows that normal CMEs have a higher ratio than inverse CMEs, due to both higher stand-off distances and lower separatrix heights. The erosion of the magnetic cloud results in a more expansion-like sheath region than inverse CMEs.

\begin{figure}[!ht]
   \centering
   \includegraphics[width=0.95\hsize]{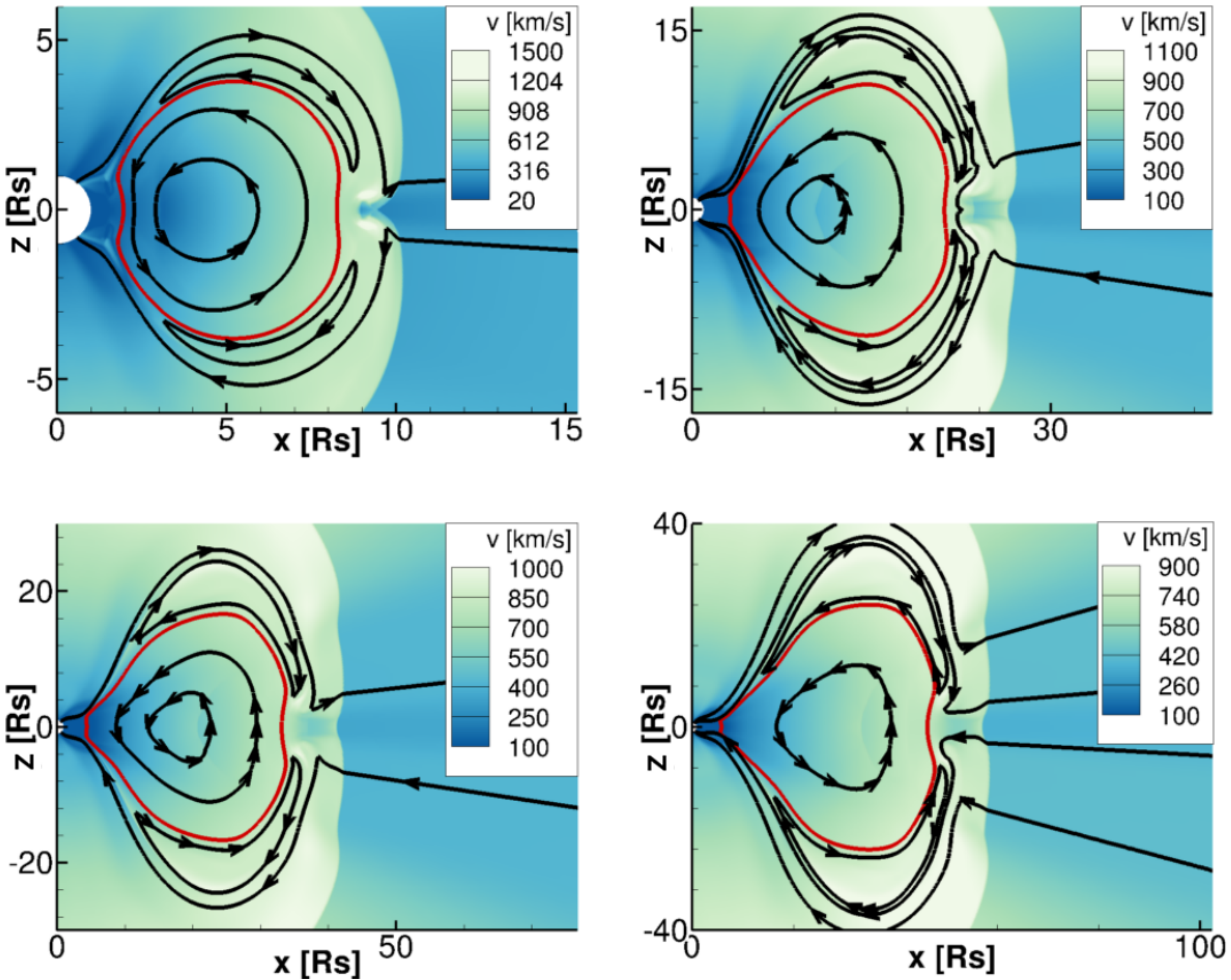}
      \caption{Snapshots of the evolution of a normal CME with an initial velocity of $400\;$km/s ejected in a low density wind, 1.5~h after ejection in the top left panel, 4.5~h in the top right panel, and 7.5~h and 11~h in the bottom left and right panels, respectively. The colours correspond to the coloured contours of the plasma velocity, the black lines represent selected magnetic field lines, and the red circle represents the edge of the magnetic cloud (i.e.\ the separatrix).
              }
         \label{fig:slownormal}
   \end{figure}

Figure~\ref{fig:slownormal} shows snapshots of the evolution of a normal slow CME in a low density wind, 1.5~h after ejection in the top left panel, 4.5~h in the top right panel, and 7.5~h and 11~h in the bottom left and right panels, respectively. The small magnetic structures at the top and bottom of the separatrix grow significantly, considerably stunting the angular growth of the magnetic cloud. Medium velocity CMEs have smaller magnetic structures so the effect on their growth is less important and fast normal CMEs do not develop them. 

\begin{figure}[!ht]
   \centering
   \includegraphics[width=0.95\hsize]{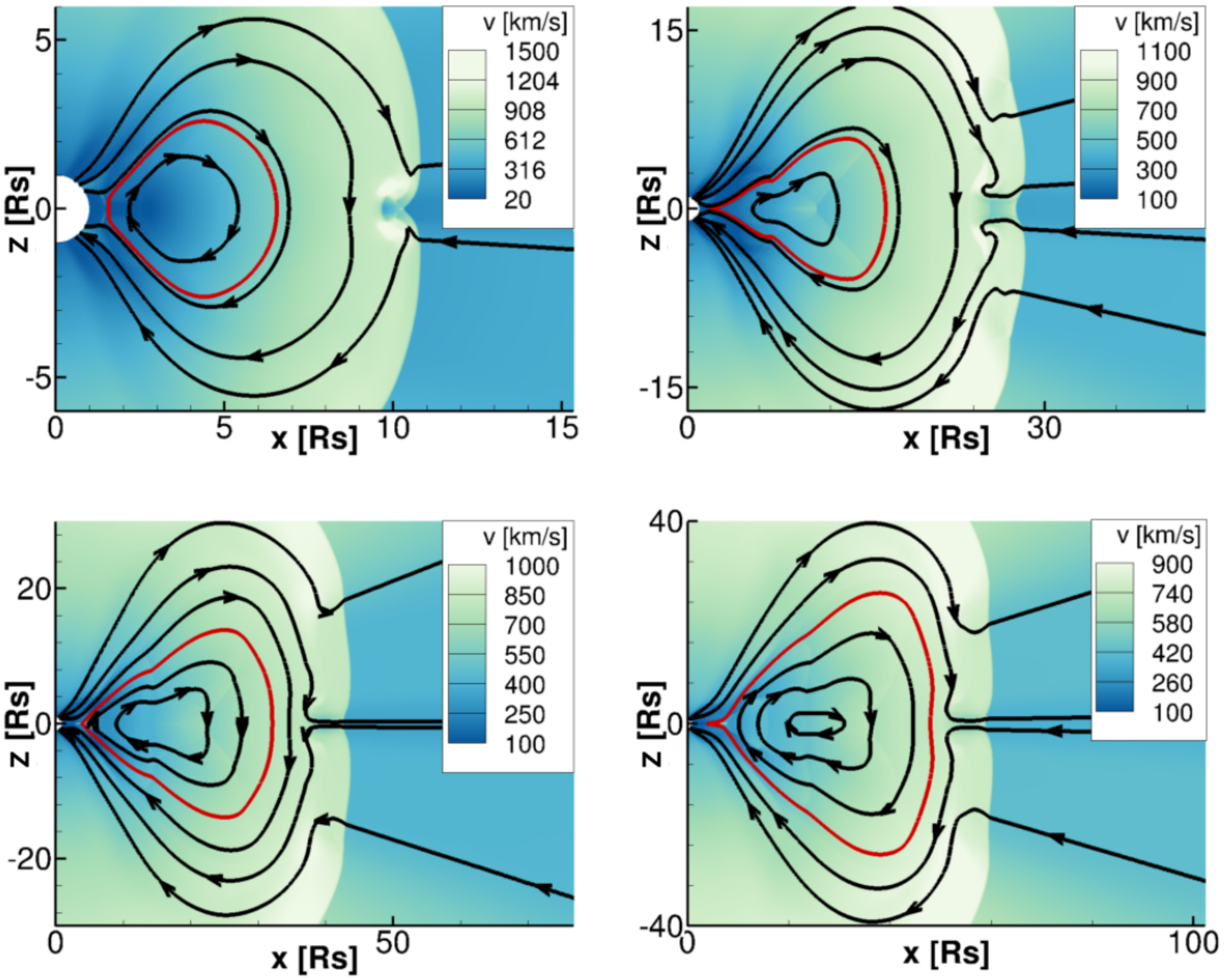}
      \caption{Snapshots of the evolution of an inverse CME with an initial velocity of $400\;$km/s ejected in a low density wind, 1.5~h after ejection in the top left panel, 4.5~h in the top right panel, and 7.5~h and 15~h in the bottom left and right panels, respectively.
              }
         \label{fig:slowinverse}
   \end{figure}

Figure~\ref{fig:slowinverse} displays similar snapshots of the progression of a slow inverse CME in a low density wind, 1.5~h after ejection in the top left panel, 4.5~h in the top right panel, and 7.5~h and 15~h in the bottom left and right panels, respectively. In the right column of Figure~\ref{fig:standoff} we can see that the properties of slow inverse CMEs show erratic behaviour before approximately 25~R$_{\odot}$. The stand-off distance actually decreases after an initial increasing phase, meaning that the separatrix travels faster than the shock. The opening angle experiences a dramatic decrease at approximately 20~$R_{\odot}$. We can see this reflected in the magnetic field lines drawn in Figure~\ref{fig:slowinverse}, where  the magnetic cloud cross-section still has a relatively circular shape (top left panel), but only three hours later (top right panel) the magnetic cloud cross-section shape displays a cusp at approximately [6,3]~$R_{\odot}$ and [6,-3]~$R_{\odot}$, giving the magnetic cloud and the field lines therein a peanut-shaped structure. Something is compressing the rear of the magnetic cloud, pinching the front of the cloud forward and deforming the field lines. Three hours later (bottom left panel) the cusps are still clearly visible in the edge of the magnetic cloud and the field lines inside. Six hours later (15 hours after ejection, bottom right panel) the cusps are much less pronounced thanks to the magnetic tension, but the field lines inside the cloud still display the peanut shaped structure. It can be argued that inverse CMEs of higher velocity also experience this pinching effect, visible as a short flattening in the beginning of the red and blue curves of bottom right panel in Figure~\ref{fig:standoff}, showing that these clouds also experience a constraining of their angular growth, but not enough to deform the cloud as much as in the slow case. \par

\begin{figure}[!ht]
   \centering
   \includegraphics[width=0.48\hsize]{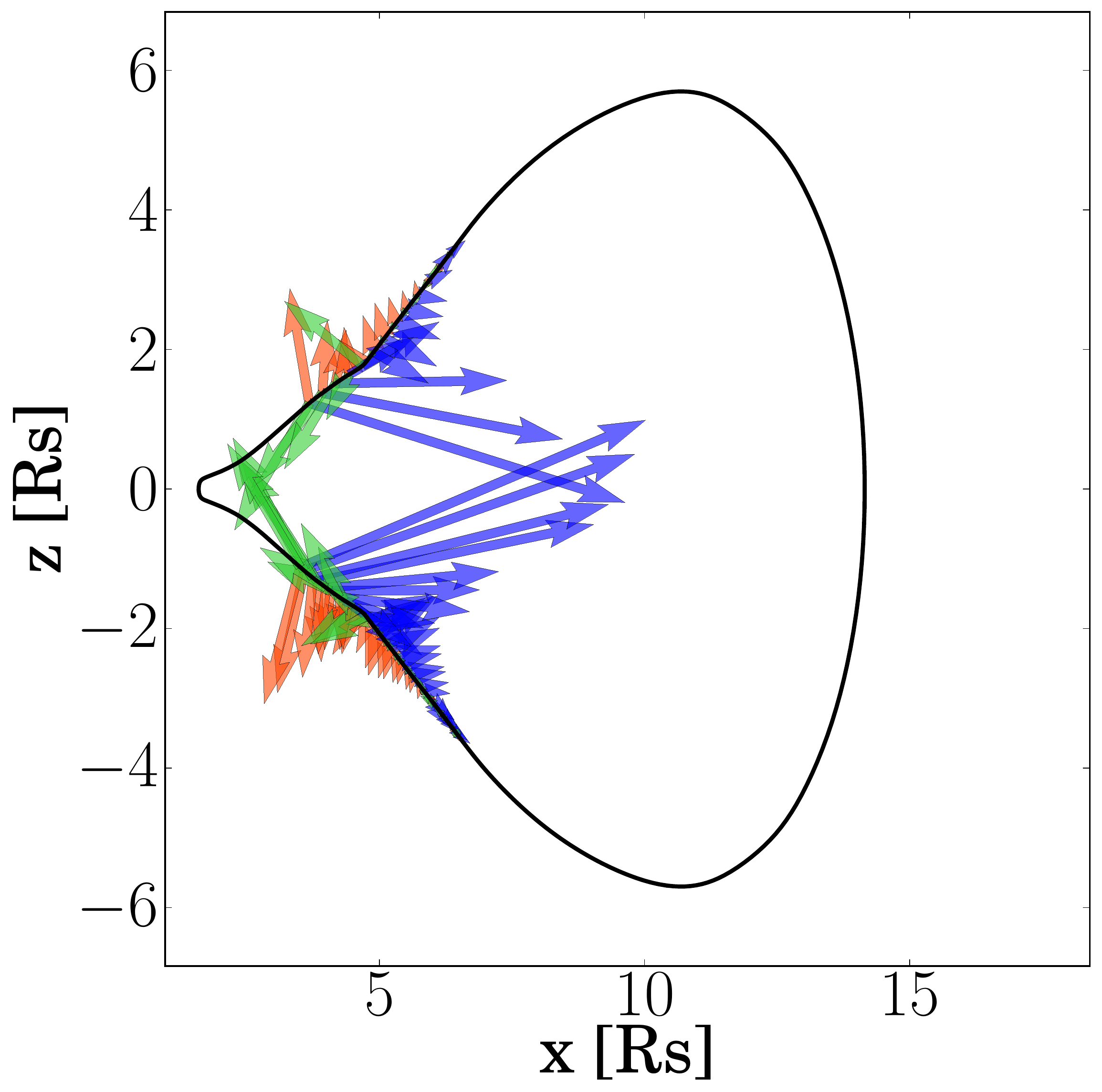}
   \includegraphics[width=0.48\hsize]{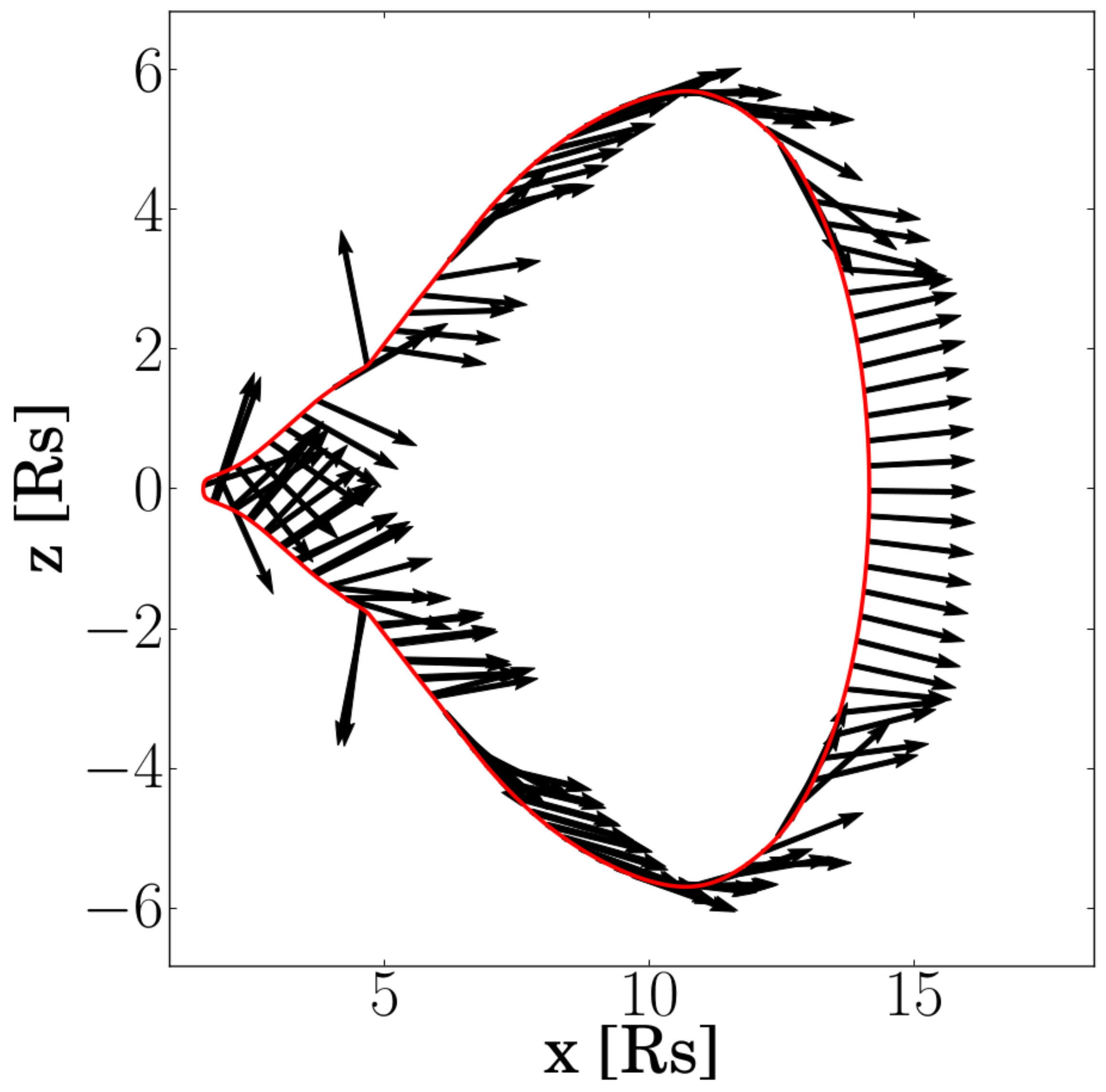}
      \caption{Left panel: Separatrix at t=3\;h for a slow inverse CME in a low density background wind, with the arrows representing the forces that working on the separatrix. The green arrows represent the magnetic tension, red arrows  the thermal pressure gradient, and blue arrows  the magnetic pressure gradient. The vectors were not normalised, so only the vectors between $x=[3.5,10]$ were drawn since that is where the deformation cusps are located and otherwise the other force vectors would obscure the view. Right panel: Total force vectors on the separatrix. Here, the vectors were normalised and drawn for the whole separatrix.
              }
         \label{fig:force_vectors}
   \end{figure}

\cite{Shen2012} performed a detailed force analysis to investigate the acceleration and deceleration of CMEs. In a similar manner, we can find the force responsible for this deformation by analysing the forces along the separatrix. The forces that govern the kinematics of a CME and that we take into consideration in our analysis are the pressure gradient, the magnetic pressure gradient, and the magnetic tension. Figure~\ref{fig:force_vectors} shows the forces individually in the left panel, while the right panel shows the total force vectors. Both panels show a clear anomaly at the location of the cusps at $\pm$[5,2]~R$_{\odot}$ and $\pm$[5,-2]~R$_{\odot}$. In the left panel we see that the force responsible for the deformation of the magnetic field lines appears to be the magnetic pressure gradient. The thermal pressure gradient expands the tail while the magnetic tension acts to `unkink' the cusps. The right panel shows that the total force is directed outwards, causing the cusps to eventually  become straight. A more detailed force analysis comparing the forces between inverse and normal CMEs will be the subject of a subsequent paper. \par

We must keep in mind that the results of the present paper are heavily dependent on our definition of the magnetic cloud as the whole region inside the separatrix and on the effects of our axisymmetric set-up. While using the separatrix gives us a mathematically justified and consistent way to study the properties of the magnetic cloud in an automated manner, it might include low-magnetic field rarefaction regions that reconnected inside the separatrix, but with distinct properties from the rest of the magnetic cloud. This effect mainly manifests itself in the extended tail region of inverse CMEs, making the length measurements of the simulations overestimations of the ejecta. Another example is the separated lobe  at the sides of normal CMEs, which in our analysis is not part of the magnetic cloud as it is a distinct magnetic structure, but contains part of the initial ejecta and shows its features in remote measurements. Another method to track the CME is by employing a relative density threshold, as done by \cite{Jacobs2005}. This would include the CME sheath as the density is increased in this region as well. We opted for our method since we focus on following the magnetic ejecta. Another alternative would be to define the CME boundary as the location where the ratio of poloidal to axial magnetic fields is highest, as done by \cite{Lugaz2013}.

\subsection{Synthetic satellite at L1}\label{subsec:sat}

\begin{figure}
\begin{center}
\includegraphics[width=0.95\linewidth]{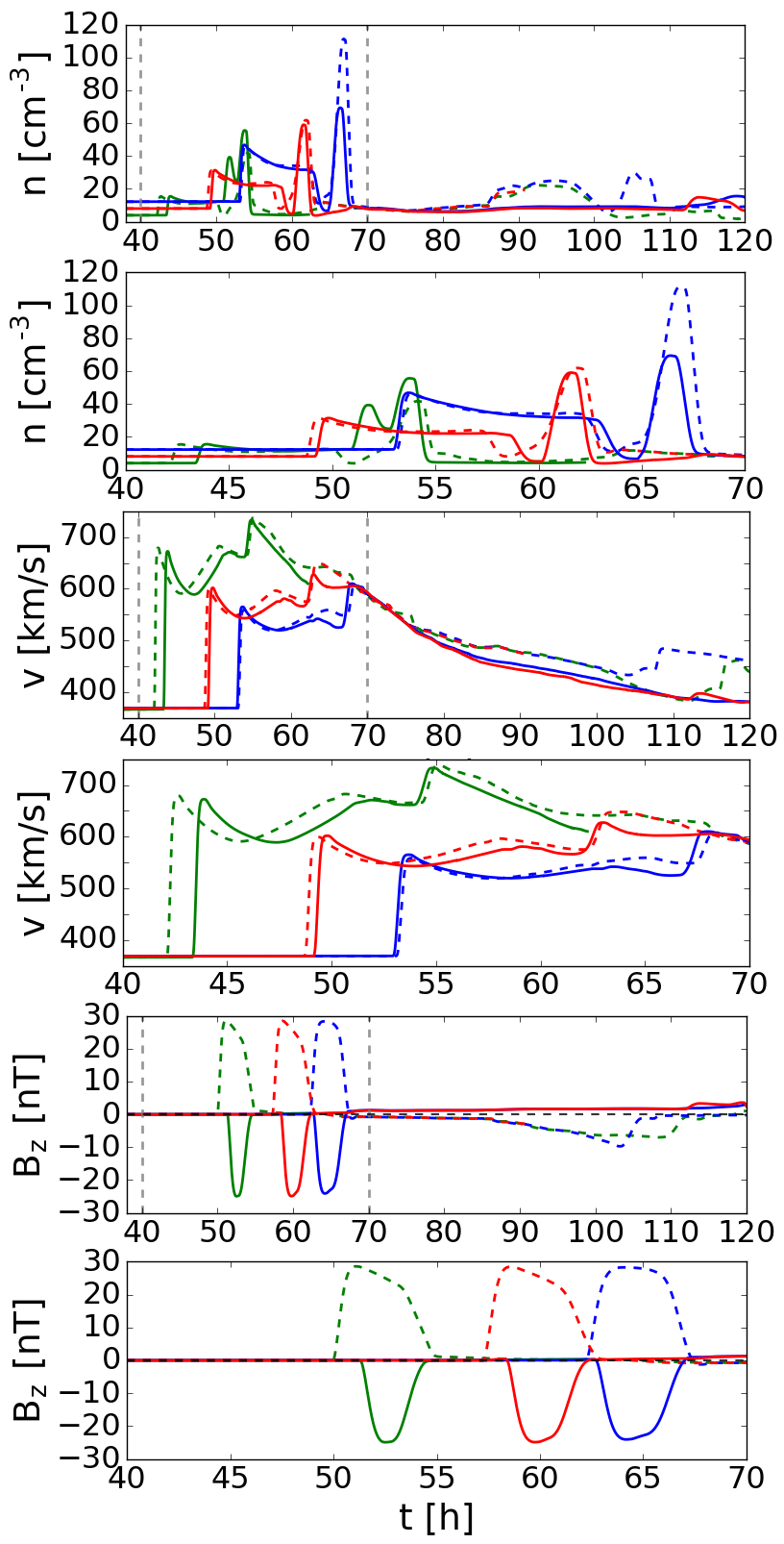}
\caption{Synthetic satellite data at L1 for ICMEs ejected with an initial velocity of $800\;$km/s with varying polarities in different background winds. The green curves represent CMEs in low density background winds, the red curves   medium density winds, and the blue curves   high density winds. Normal CMEs are shown as solid lines while inverse CMEs are shown as dashed lines. The black dashed line in the two lowest plots represents $B_z=0$. Each second plot is a zoom-in on the marked region between grey dashed lines in the plot above.}\label{fig:sat}
\end{center}
\end{figure}

By measuring the temporal evolution in a chosen grid cell, we can synthesise satellite measurements of the simulated ICMEs at any radial distance or latitude. We have determined and compared synthetic satellite measurements at L1, which is at about 212 $R_{\odot}$, to compare the observational properties of the different simulations. Figure~\ref{fig:sat} shows the synthetic satellite data of the simulations with an initial velocity of $800\;$km/s.

The two upper  panels show the measured number density, where the first panel shows the whole ejecta passing the spacecraft and the second panel only shows the sheath and front of the magnetic cloud. A similar structure to the radial cross section of the number density in Figure~\ref{fig:cross_section} can be recognised in the second panel. We can see that all ejecta follow a comparable profile, so the time stamps mentioned in the following description will refer to the normal $800\;$km/s CME in a medium density background wind. The shock can be identified as a sharp density increase at approximately 48~h. The magnitude of the density jump is approximately the same for normal and inverse CMEs and increases for increasing background wind density, as more interplanetary material accumulates at the shock. The sheath region follows the shock  until approximately 58~h, where the density increase for increasing background wind density can also be seen. After the sheath the front of the separatrix can be distinguished as a dip in density, where the high magnetic pressure gradient seems to have displaced to plasma. The dip is followed by a density peak, where the plasma is compressed  between the separatrix and the centre of the cloud. It is noteworthy that the density dip for higher density winds is more severe relative to the density of the preceding sheath and of the following peak. In contrast to the rest of the eruption, we find that the magnitude of the density peak inside the magnetic cloud differs between normal and inverse CMEs. The density peak for inverse CMEs is lower, approximately the same and much higher with respect to normal CMEs for $4\;$cm\textsuperscript{-3}, $8\;$cm\textsuperscript{-3}, and $12\;$cm\textsuperscript{-3} solar wind densities, respectively. Especially for our high density wind, the artificial satellite measurements of the density peak are higher than observed number density values, which are closer to 40\;cm\textsuperscript{-3}, for example seen in the April 2000 CME studied by \cite{Chane2008}. This is possibly due to the combination of having either a very high initial density eruption and a high density of the wind itself in case of the blue curve. This is an artefact of our idealised 2.5D set-up. At approximately 63~h, following the peak, there is a region of low density due to the CME expansion. The top panel also shows additional density structures for inverse CME measurements. Reconnection in the tail of the magnetic cloud forms small magnetic islands, which propagate behind the CME. 
 
The two central  panels of Figure~\ref{fig:sat} show the plasma speeds. The plasma speed jumps at the shock from a background wind speed of approximately $370\;$km/s to $679\;$km/s and $672\;$km/s for the inverse and normal CMEs in low density winds, respectively; to $599\;$km/s and $602\;$km/s for the medium density winds; and $558\;$km/s and $565\;$km/s for the high density winds. The velocity jump magnitude increasing and the density jump decreasing for decreasing solar wind densities is consistent with the conservation of mass. Big jumps in densities imply small jumps in shock normal speed.  The plasma speed in the sheath decreases initially, but steadily increases again until the front of the separatrix is measured at approximately 58~h. Between 58~h and 62~h there is a small dip in plasma speed in the front part of the magnetic cloud, coinciding with the density peak in the upper panel. After this dip there is an increase of the plasma speed as the centre of the magnetic cloud crosses the artificial satellite. The time series all follow a similar profile for different CME polarities and since increasing the background wind density increases the drag, we then find lower plasma speed and later arrival times.

Finally, the two bottom  panels display the evolution of $B_z$ at L1, where naturally the $B_z$ for an inverse and normal CME have opposite signs. Since we are observing on the equatorial plane, the shock cannot be distinguished in these profiles and the value of $B_z$ only deviates from zero close to the separatrix (58~h). Then it reaches a peak value between the sheath and the centre of the magnetic cloud, as the magnetic field lines are compressed significantly in this region. Furthermore, the $B_z$ peak is wider and  stronger for inverse CMEs, as the distance between the sheath and the centre of the magnetic cloud is larger for inverse CMEs and normal CMEs erode at the front due to magnetic reconnection. The background wind density affects the arrival time but has very little effect on the magnitude of the $B_z$ component. Only normal CMEs will be able to cause geomagnetic storms since only they have a strong negative $B_z$, allowing for reconnection with the magnetosphere of the Earth. The fifth panel of Figure \ref{fig:sat} shows that although signatures of a fully rotating $B_z$ are visible, there is a clear asymmetry in the north--south magnetic field. The $B_z$ magnitude is much weaker behind the centre of the magnetic cloud (at approximately 65~h, where $B_z$ changes sign) compared to the front of the magnetic cloud. It is also much weaker than measured values in real observations. Our extremely high initial magnetic cloud momentum and pressure leads to our CMEs experiencing severe and asymmetric expansion. Therefore, the magnetic field lines are strongly compressed at the front of the magnetic cloud, while the field becomes stretched at the rear, leading to a stronger $B_z$ at the front and a weaker $B_z$ at the rear. In simulations performed with a lower initial pressure and lower momentum, as done in \cite{Chane2008}, artificial satellite data shows comparable plasma and magnetic field signatures but a stronger $B_z$ magnitude is visible at the back of the magnetic cloud.

\begin{figure}
\begin{center}
\includegraphics[width=0.95\linewidth]{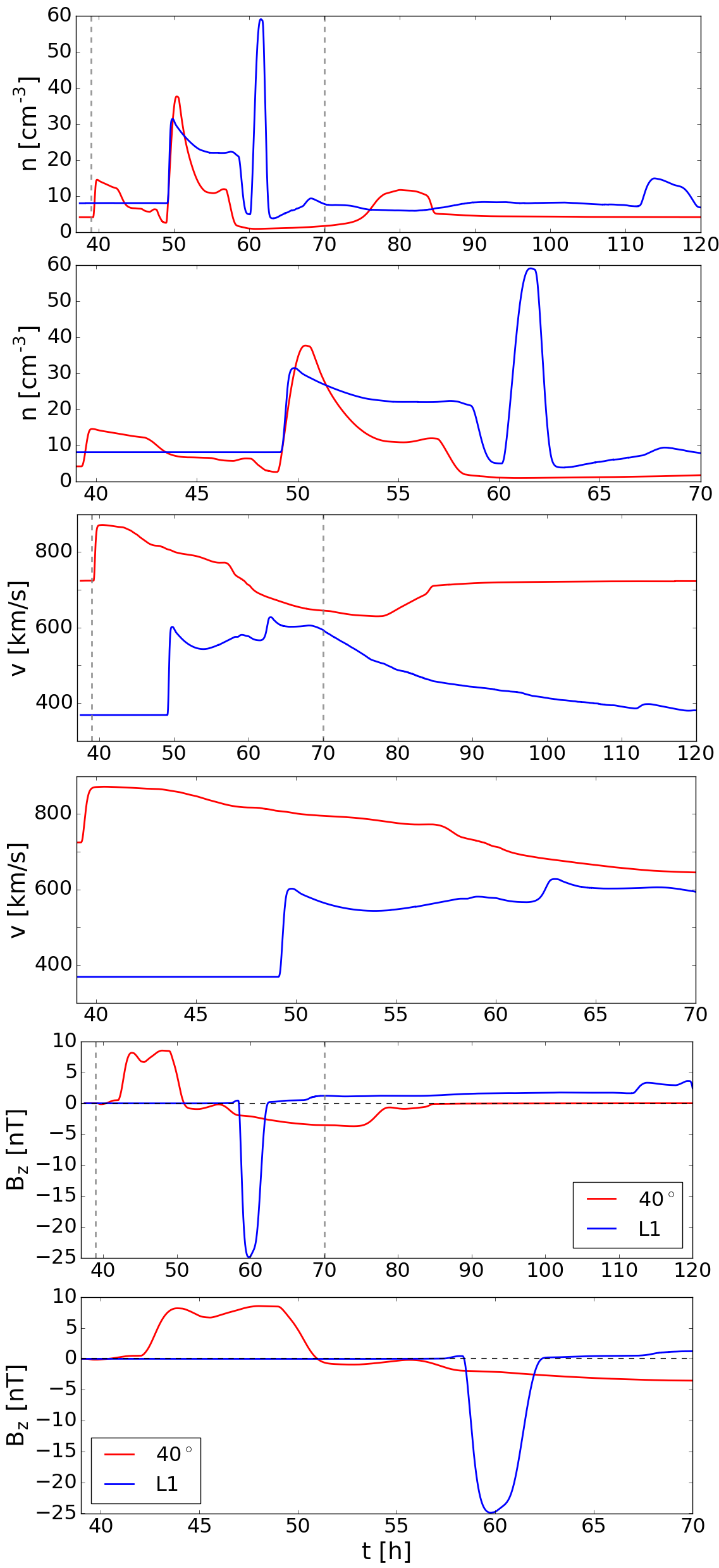}
\caption{Synthetic satellite data at L1 and at a 40\degree\ angle for a normal CME ejected with an initial velocity of $800\;$km/s in a $8\;$cm\textsuperscript{-3} density background wind. Each second plot is a zoom-in on the marked region between grey dashed lines in the plot above. The black dashed line in the two lowest plots represents $B_z=0$.}\label{fig:degsat}
\end{center}
\end{figure}

In Figure~\ref{fig:together}, a closed magnetic structure of significant size has formed above and below the separatrix (at approximately $[120,100]\;R_{\odot}$ and $[120,-100]\;R_{\odot}$ in the rightmost panel of the top row), with opposite polarity to that of the magnetic cloud itself. By comparing synthetic satellite data at L1 and that of a satellite placed at the same radial distance but at approximately 40$\degree$ latitude, we can compare the observational signatures of the magnetic cloud and the other magnetic structure. A latitude of 40$\degree$ falls just outside of the separatrix (red line in Figure~\ref{fig:together}). The red curve in Figure~\ref{fig:degsat} shows synthetic satellite data at higher latitudes, while the blue curve shows data at L1. The magnetic structure at higher latitude is significantly faster than the magnetic cloud due to the solar wind speed being higher while its density is lower at higher latitudes, thus reducing the drag. We see in the upper panel that the smaller structure has lower density, but the same density profile as the magnetic cloud. It is also preceded by a shock, with a compression ratio of 3.5, has a sheath region of higher density which is then followed by a density dip and density peak. The central panel shows that the jump in solar wind speed is lower for the smaller magnetic structure ($148\;$km/s compared to $216\;$km/s for the magnetic cloud) and that there is no acceleration of the plasma as the cloud crosses the satellite. The bottom panel shows that the smaller structure has an smaller absolute value of $B_z$ and is of opposite polarity. The polarity of the magnetic field being opposite to that of the magnetic cloud might be a reason to exercise caution when interpreting in~situ CME observations, especially when it is single-spacecraft. Measuring the $B_z$ value of only the small structure would imply that the ejection is has little or no geo-effectiveness. An in~situ observation at a lower latitude indicates the opposite, that  the ejection is very geo-effective thanks to its large negative $B_z$. The smaller magnetic structures thus may have a large influence on identifying and forecast the effect of ICMEs, unless measurements at multiple latitudes are made simultaneously. \par

These small-scale structures are also called magnetic islands and are formed by tearing instabilities in current sheets. They have been observed in~situ by  \cite{Chian2011}, observing magnetic islands at the front of an ICME, and \cite{Eriksson2014}. To resolve these tearing instabilities and their resulting magnetic islands in numerical simulations, a high resolution at the current sheets is necessary. Recent progress in this area has been made by \cite{Karpen2012}, \cite{Guidoni2016} and \cite{Hosteaux2018}. They are not present at the front of an inverse CME, but at the rear because this is where a current sheet forms when the CME has the same polarity as the solar wind.

\section{Conclusions}
In an attempt to quantify the effects of the polarity of the internal magnetic field of CMEs and of the density of the background solar wind on the evolution of the CMEs from the solar corona to the Earth, we set up some numerical experiments. We used the MPI-AMRVAC code in spherical coordinates on a stretched grid with four levels of refinement. We considered an axisymmetric (2.5D) set-up, but made sure the results of the simulations   mimic  fully 3D simulations \citep{Jacobs2007}. We first modelled the background solar wind under solar minimum conditions and considered steady bi-modal wind solutions with three different densities, namely a density of 4 cm\textsuperscript{-3} on the equatorial plane at 1~AU, another wind with a density of 8 cm\textsuperscript{-3} and one with a density of 12 cm\textsuperscript{-3}, which were referred to as the low, medium and high density solar wind, respectively. In order to keep full control of the initial CME parameters, we omitted the initialisation phase of the CMEs and focused on their evolution. We thus modelled the CMEs as magnetised plasma blobs with a higher pressure and density than the surrounding coronal plasma and with a smooth initial velocity profile ($400\;$km/s, $800\;$km/s and $1200\;$km/s) and two different magnetic field polarities (normal and inverse), as did \cite{Chane2006}. We simulated the interaction of the magnetised plasma blobs with the surrounding solar wind, focusing on the kinematics of the magnetic clouds, the shock waves they generate, and their deformation and erosion, all the way from the corona to their arrival at Earth (i.e.\ 1~AU).\par

We quantified the effects of the polarity of the internal magnetic field of the CMEs and of the density of the background solar wind on the arrival times of the shock front and the magnetic cloud. The magnetic cloud was determined as the volume within the separatrix. This clear definition enabled us to determine the positions and propagation velocities of the centre and the edges of the magnetic clouds and thus also the stand-off distance of the leading shock fronts (i.e.\ the thickness of the magnetic sheath region) and the deformation and erosion of the magnetic clouds during their evolution from the Sun to the Earth. We found that lowering the background density or increasing the initial CME velocity lowers the arrival time of the CME. Table \ref{table:arrival_diff} shows the arrival time difference for the magnetic clouds of CMEs ejected in the same initial conditions, except their polarity differs. Inverse CMEs arrive faster than normal CMEs for all background wind densities and initial velocities assumed in this paper. The lowest wind density and lowest initial velocity used in our simulation yield an arrival time of 4.5~hours earlier for inverse CMEs. Increasing the background wind density or increasing the initial velocity leads to a lower arrival time difference.

\begin{table}[htb]
\centering
\begin{tabular}{|c|c|c|c|}
\hline 
  & 400 km/s & 800 km/s & 1200 km/s \\
\hline 
4 cm\textsuperscript{-3} & 4.5 & 1.5 & - \\
\hline 
8 cm\textsuperscript{-3} & 1.5 & 1 & 1 \\
\hline 
12 cm\textsuperscript{-3} & 1.5 & 1 & 0.5 \\
\hline 
\end{tabular}  
 \caption{Difference in arrival time of the magnetic cloud at 1~AU between normal and inverse CMEs expressed in hours after ejection. Positive values mean inverse CMEs arrive faster, while negative values mean that normal CMEs arrive faster.}\label{table:arrival_diff}
\end{table}

The polarity of the internal magnetic field of the CME has a substantial effect on its propagation velocity and on its deformation and erosion during its evolution towards Earth. The initial acceleration of normal CMEs is higher than that of inverse CMEs, but inverse CMEs overtake their normal counterparts when still close to solar surface ($\pm$25~R$_{\odot}$). Higher background wind densities and higher initial velocities lead to a stronger deceleration of the CME due to an increased drag effect.

\begin{table}[]
\begin{tabular}{cccc}
\hline
\multicolumn{1}{|c|}{\textbf{MC length}}     & \multicolumn{1}{c|}{400~km/s} & \multicolumn{1}{c|}{800~km/s} & \multicolumn{1}{c|}{1200~km/s} \\ \hline
\multicolumn{1}{|c|}{4~$\rm cm^{-3}$}                   & \multicolumn{1}{c|}{-23}      & \multicolumn{1}{c|}{-48}      & \multicolumn{1}{c|}{-}         \\ \hline
\multicolumn{1}{|c|}{8~$\rm cm^{-3}$}                   & \multicolumn{1}{c|}{-37}      & \multicolumn{1}{c|}{-55}      & \multicolumn{1}{c|}{-82}       \\ \hline
\multicolumn{1}{|c|}{12~$\rm cm^{-3}$}                  & \multicolumn{1}{c|}{-59}      & \multicolumn{1}{c|}{-73}      & \multicolumn{1}{c|}{-93}       \\ \hline
\multicolumn{1}{l}{}                         & \multicolumn{1}{l}{}          & \multicolumn{1}{l}{}          & \multicolumn{1}{l}{}           \\ \hline
\multicolumn{1}{|c|}{\textbf{Opening angle}} & \multicolumn{1}{c|}{400~km/s} & \multicolumn{1}{c|}{800~km/s} & \multicolumn{1}{c|}{1200~km/s} \\ \hline
\multicolumn{1}{|c|}{4~$\rm cm^{-3}$}                   & \multicolumn{1}{c|}{-8}       & \multicolumn{1}{c|}{-16}      & \multicolumn{1}{c|}{-}         \\ \hline
\multicolumn{1}{|c|}{8~$\rm cm^{-3}$}                   & \multicolumn{1}{c|}{-7}       & \multicolumn{1}{c|}{-14}      & \multicolumn{1}{c|}{-17}       \\ \hline
\multicolumn{1}{|c|}{12~$\rm cm^{-3}$}                  & \multicolumn{1}{c|}{-5}       & \multicolumn{1}{c|}{-11}      & \multicolumn{1}{c|}{-12}       \\ \hline
\multicolumn{1}{l}{}                         & \multicolumn{1}{l}{}          & \multicolumn{1}{l}{}          & \multicolumn{1}{l}{}           \\ \hline
\multicolumn{1}{|c|}{\textbf{Stand-off}}      & \multicolumn{1}{c|}{400~km/s} & \multicolumn{1}{c|}{800~km/s} & \multicolumn{1}{c|}{1200~km/s} \\ \hline
\multicolumn{1}{|c|}{4~$\rm cm^{-3}$}                   & \multicolumn{1}{c|}{5.2}      & \multicolumn{1}{c|}{1.7}      & \multicolumn{1}{c|}{-}         \\ \hline
\multicolumn{1}{|c|}{8~$\rm cm^{-3}$}                   & \multicolumn{1}{c|}{5.1}      & \multicolumn{1}{c|}{2.5}      & \multicolumn{1}{c|}{6.5}       \\ \hline
\multicolumn{1}{|c|}{12~$\rm cm^{-3}$}                  & \multicolumn{1}{c|}{4.2}      & \multicolumn{1}{c|}{2.9}      & \multicolumn{1}{c|}{6.4}       \\ \hline
\end{tabular}
\caption{Upper table: Difference in separatrix length between normal and inverse CMEs in R$_{\odot}$ for the same initial CME velocity and background wind density at 1~AU. Middle and lower tables: Similar values, but for the opening angle (in degrees) and stand-off distance (in R$_{\odot}$), respectively. Negative values mean that the property for the inverse case is larger than that of the normal case.}
\end{table}

Shocks of inverse CMEs initially have higher stand-off distances than their normal counterparts, but after approximately 14~R$_{\odot}$ the stand-off distances of all inverse CME shocks became lower than those of normal CMEs under the same conditions. At the end of the simulation, the difference between the stand-off distance for inverse CME shocks ranges from approximately 2~R$_{\odot}$ for fast CMEs to approximately 6~R$_{\odot}$ for medium velocity CMEs. The initial speed of the ICME and the solar wind density have little influence on the stand-off distance for CMEs with the same polarity, except in the case of a low density background wind where the effect of the drag is smaller compared to the other simulations. In addition, normal CMEs ejected in the same background wind with a higher initial velocity are shorter when their magnetic cloud reaches the same distance from the Sun, and CMEs with the same velocity in a higher density wind experience a stronger drag than those ejected in lower density winds, causing longer travel times, but at the same time the magnetic cloud also becomes more compressed. Inverse CMEs are much more elongated in the radial direction compared to normal CMEs. The effect of the polarity on the length of the separatrix increases for higher background wind density and higher initial CME velocity. Our high initial pressure leads to lengths of approximately two times the values found in observations.
Finally, inverse CMEs were found to have a larger opening angle, with the difference in opening angle increasing for lower background wind densities and higher initial CME velocities. Keeping the same polarity and background wind density, the opening angle increases when the initial velocity is higher for the medium and high velocity CMEs. The lower velocity CMEs experience a different expansion compared to the other velocities. In addition, similar  to the effect on the CME radial length, lowering the background wind density for both normal and inverse CMEs with the same initial speed results in larger opening angles.

The evolution of CMEs with opposite polarities differs due to their reconnection processes. Normal CMEs have the opposite magnetic polarity from the background wind, which implies that magnetic reconnection occurs at their fronts. The polarity of the magnetic field of an inverse CME is the same as that of the surrounding wind and magnetic reconnection occurs behind  these  CMEs.

Inverse CMEs also have substantially more massive magnetic clouds than normal CMEs that are launched with the same initial velocity and ejected in the same background wind. When magnetic reconnection opens up the field lines of the helmet streamer surrounding the plasma blob, reconnection at the rear enables more magnetic flux to be added to the magnetic cloud compared to reconnection at the front in the normal CME case. 

By measuring the temporal evolution at about 212 $R_{\odot}$, we synthesised satellite measurements of the simulated ICMEs at L1. We found that the structure and arrival time of normal and inverse CMEs were comparable. A different polarity has little effect on the strength of the shock. However, the density peak of the magnetic cloud changes dramatically. For normal CMEs the peak density is approximately 60~$\rm cm^{-3}$ for all initial conditions, while for inverse CMEs the peak density rises to approximately 40, 60 and 110~$\mathrm{cm^{-3}}$ for low, medium, and high background wind densities, respectively. Increasing the background wind density also lowers the velocity jump magnitude at the shock. Finally, we also quantified the evolution of the $B_z$ component at L1, where naturally the $B_z$ components of the inverse and normal CMEs have opposite signs. The value of $B_z$ reaches a peak value between the separatrix front and the centre of the magnetic cloud, as the magnetic field lines are compressed significantly in this region. This $B_z$ peak is wider and stronger for inverse CMEs, but the background wind density has, except for the arrival time, very little effect on the $B_z$ component. For a background wind polarity that was used in our simulations, only normal CMEs are able to cause geomagnetic storms since they have a strong negative $B_z$, allowing for reconnection with the magnetosphere of the Earth. If the solar dipole field switches for a different solar cycle, CMEs with an inverse polarity would be geoffective. We found that normal CMEs develop magnetic structures at the top and bottom of the separatrix. Synthetic satellite data showed that measuring at a different latitude dramatically influences the conclusions of the $B_z$ data.

\begin{acknowledgements}
EC was funded by the Research Foundation - Flanders (grant FWO 12M0119N).
These results were obtained in the framework of the projects
GOA/2015-014 (KU Leuven), G.0A23.16N (FWO-Vlaanderen) and C~90347 (ESA Prodex).
For the computations we used the infrastructure of the VSC – Flemish
Supercomputer Center, funded by the Hercules foundation and the Flemish
Government – department EWI.
\end{acknowledgements}

\bibliography{bib}
\bibliographystyle{aa}

\end{document}